\documentclass[a4paper,11pt]{article}
\usepackage{jcappub} 
\usepackage{multicol}
\usepackage{subcaption}
\usepackage{orcidlink} 
\usepackage{lineno}
\usepackage{hyperref}
\usepackage{subcaption}
\usepackage{xcolor}
\usepackage{soul}
\usepackage{multirow}
\usepackage{booktabs}
\usepackage{tikz}

\usetikzlibrary{positioning,arrows.meta,shapes.geometric,positioning}
\title{\boldmath On the bin sensitivity of the transverse BAO}

\tikzset{
    startstop/.style = {rectangle, rounded corners, minimum width=2.0cm, minimum height=0.6cm, text centered, draw=black, fill=blue!10, font=\footnotesize},
    process/.style   = {rectangle, minimum width=2.0cm, minimum height=0.6cm, text centered, draw=black, fill=cyan!10, font=\footnotesize},
    decision/.style  = {diamond, aspect=1.2, minimum width=1.8cm, inner sep=1pt, text centered, draw=black, fill=gray!10, font=\footnotesize},
    arrow/.style     = {thick, -{Stealth[scale=0.6]}}
}
\author[a,b,c]{Paula S. Ferreira\orcidlink{0000-0002-7540-040X},}

\affiliation[a]{Observatório Nacional,\\
 Rua General José Cristino, 77, São Cristóvão, Rio de Janeiro-RJ. CEP 20921-400, Brazil}
 
 \affiliation[b]{Center for Cosmology and Computational Astrophysics, \\ Institute for Advanced Study in Physics, Zhejiang University, Hangzhou 310058, China.}
\affiliation[c]{Institute of Astronomy, School of Physics, Zhejiang University, Hangzhou 310058, China}
\author[a]{Carlos A. P. Bengaly \orcidlink{0000-0001-5731-3348}}

\emailAdd{paulaferreira@on.br}

\abstract{The BAO characteristic scale is a useful tool for understanding the evolution of the universe, especially the influence of dark energy on this evolution. In this work, we study the projection effect in transverse BAO, namely the mixing of BAO signals from different epochs caused by $z_c$ uncertainty within a chosen bin. We focus our forecast on two surveys of interest: the Square Kilometre Array (SKA) HI galaxy redshift survey and the Dark Energy Spectroscopic Instrument (DESI)'s final Luminous Red Galaxy (LRG) sample. We test the sensitivity in finding the transverse BAO depending of three bin configurations: a Gaussian, a top-hat and an intermediate of them semi-Gaussian. We also analyse the precision the bin widths $\sigma_z$ from smaller to wider bins $0.01<\sigma_z<0.1$. In order to correct these deviations, we propose a correction based on adjacent redshift to $z_c$, this would provide a statistical correction instead of only relying on  fiducial cosmology.
Finally, we conclude that despite the higher shot-noise than the top-hat bin separation, the semi-Gaussian bin is the most accurate case to find the BAO signal for most bin widths ($\sigma_z<0.04$) we tested. For a SKA-like survey, Gaussian binning yields the weakest parameter constraints among the three options. In contrast, for a DESI-like survey, this trend reverses: the semi-Gaussian scheme provides the tightest constraints, but only when $\sigma_z<0.04$, while for larger $\sigma_z$ values the Gaussian and top-hat binnings deliver similarly strong constraints.  For a real data pipeline, one must choose which approach is best: well constrained cosmological parameters, but less accurate BAO measurement; or well measured BAO peak/wiggles. In case one is interested in combining the BAO measurement with other cosmological probes, it might be more interesting to focus on the bin scheme that is statistically robust and with a sharp representation of the redshift of interest, which is the Semi-Gaussian case. }

\begin{document}
\maketitle
\flushbottom

\section{Introduction}

Based on the standard cosmological model, $\Lambda$CDM, the large-scale structure (LSS) formed due to the growth of density fluctuations after inflation. During the epoch up to redshift $z\simeq 1000$ baryon and photons were tightly coupled in a fluid propagating sound waves that formed thanks to the competition between baryon's gravity and photon's radiation pressure. Around $z\simeq1100$ the universe cooled and the coupling ended leaving a signal in the LSS of the last spherical wave-front from the oscillations with a comoving radius of $r_s\simeq 150$ Mpc which is imprinted as a preferred scale for galaxies to cluster, this probe is known as Baryon Acoustic Oscillations (BAO) \cite{Peebles1970,Sunyaev1970}. The BAO characteristic scale is a useful tool for understanding the evolution of the universe, especially the influence of dark energy on this evolution.

Due to the relation with the clustering of galaxies, we rely on statistical tools to find the BAO feature. One is the two-point correlation function  that measures the excess probability, over a random (unclustered) distribution, of finding two objects at a given angular ($\theta$) or radial ($r$) separation \cite{peebles1973statistical}. The other is its Fourier transform, the power spectrum, also can be either angular ($C_\ell$) or in real space 3D ($P(k)$). 

The first spectroscopic survey detections in galaxy clustering were made almost simultaneously by \cite{Eisenstein2005} and \cite{Cole2005}. Recent results using the three-dimensional analysis from the Sloan Digital Sky Survey (SDSS)'s extended Baryon Oscillation Spectroscopic Survey (BOSS) \cite{Anderson2012,Alam2017} and Extended Baryon Oscillation Spectroscopic Survey (eBOSS) \cite{eBOSS2020} and the second year of Dark Energy Spectroscopic Instrument (DESI) \cite{desicollaboration2025datarelease1dark} now provide percent-level precision on the cosmic distance scale across a wide range of redshift. 

Due to the lower resolution, it is preferable to use photometric surveys to measure the BAO in angular scale ($\theta_{\rm BAO}$), as this prevents the smearing of the signal along the line of sight. The main results of these types of experiments came from the Dark Energy Survey (DES) \cite{menafernández2024darkenergysurveygalaxy}. 

Although the three-dimensional correlation function or power spectrum can be measured with high precision, using only transverse information offers a valuable way to reduce reliance on an assumed fiducial cosmological model. This can be implemented, for instance, through the parametrised framework proposed in \cite{Sanchez_2010}. Such a methodology has been applied in several studies, initially by \cite{carvalho2016baryon} and subsequently to a low-redshift sample by \cite{alcaniz2016measuringbaryonacousticoscillations}. The following papers were then a continuation of the methodology used: \cite{de2018angular,carvalho2020transverse,de2021bao,de2020baryon,menote2022baryon,Ferreira2024,ferreira2025angularbaomeasurementsdesi}. Additionally, in \cite{marra2019first}, the authors measured the BAO radial feature using the same polynomial function but applied to the redshift space.

\begin{figure}[ht]
    \centering
    \includegraphics[width=1\linewidth]{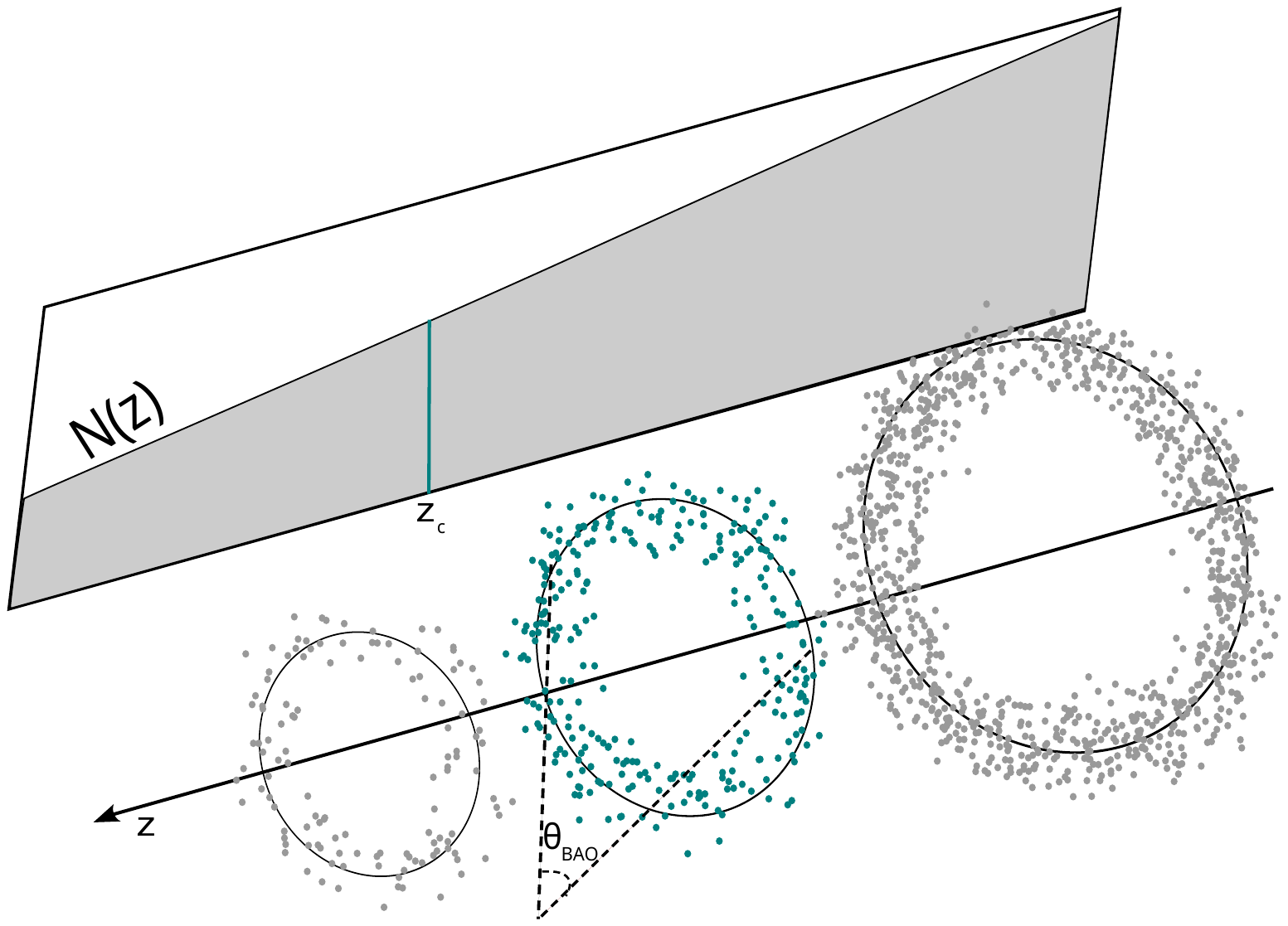}
    \caption{Representation of the projection effect in the transverse BAO feature. The z axis points to increasing redshift, when the BAO feature has a reduced angular size. The circles represent the expected BAO projected in the sky, with a radius of $\theta_{\rm BAO}$, the dots can be thought as detected galaxies distributed preferably at this scale. A background projection shows the bin redshift distribution $N(z)$, in this example, the central redshift $z_c$ is the desired redshift for the BAO feature we are looking for, but the projection effect due to the galaxies in the denser and less dense regions can dilute the signal.}
    \label{fig:proj_effect}
\end{figure}

During these years, the groups involved in detecting $\theta_{\rm BAO}$ have used either a Gaussian or a top-hat redshift bin separation. However, no one has considered the implications of the bin choice in BAO detection. One problem may introduce spurious $\theta_{\rm BAO}$ results: the projection effect. In \cite{Sanchez_2010}, it was found that the projection effect increases with bin width and decreases with the bins' central redshift, $z_c$. This effect is the mixing of BAO signals from different epochs due to $z_c$ uncertainty in a chosen bin. This is summarised in Figure~\ref{fig:proj_effect}, where the z-axis indicates an increasing redshift, where the angular size of the BAO feature reduces. The circles show the expected BAO projected in the sky with radius $\theta_{\rm BAO}$. Dots represent galaxies, preferably distributed at this scale. A background projection shows the bin redshift distribution $N(z)$, with $z_c$ as the target for the BAO feature. The illustration highlights the projection effect: we aim to determine $\theta_{\rm BAO}$ at $z_c$. Ideally, the bin would be represented by a systematic-weighted mean redshift based on the survey configuration, but additional epochs within the bin weaken the signal.

In addition to affecting all redshift ranges, the projection effect is more problematic for low redshift samples. The main problem with these samples is the redshift space distortions (RSD), a galaxy's observed redshift is a combination of the cosmological expansion (Hubble flow) and its peculiar velocity (motion due to local gravitational fields). For the transverse BAO, the Fingers-of-God Effect (FoG) \cite{Hamilton_1998,kaiser87} is the main distortion. Galaxy clusters exhibit large velocity dispersion due to random, virialised motions, elongating the cluster along the line-of-sight when converting redshifts into distances. This elongation smears the galaxy distribution, projecting galaxies from different radial distances into the same transverse bin, blurring the BAO ring. At low redshifts ($z < 0.5$), gravitational collapse results in more massive, virialised clusters and a more non-linear cosmic web, intensifying FoG with larger random velocities and smearing scale. 

In the coming years, surveys such as the Legacy Survey of Space and Time (LSST) \cite{collaboration2012large}, the \textit{Euclid} Wide Survey \cite{scaramella2022euclid}, the \textit{Roman Space Telescope} \cite{eifler2021cosmology}, and the \textit{China's Space Survey Telescope} (CSST) survey \cite{zhan2021wide,miao2024forecasting}, are expected to enhance the detection of a greater number of objects, identified through either spectroscopic or photometric redshifts. Therefore, the vast number of observed galaxies is susceptible to systematics and the projection effect. Usually, big collaborations take advantage of the Reconstruction \cite{Eisenstein_2007} to remove non-linearity effects, but with the price of introducing a fiducial cosmology in the analysis.

In this work, we try to understand, through Fisher Forecast, the impact of the choice of bin separation on the finding of $\theta_{\rm BAO}$. We conducted the analysis for two different surveys of interest: the Square Kilometre Array (SKA) HI galaxy redshift survey and DESI's final Luminous Red Galaxy (LRG) sample. In particular, we test three types of bins: a Gaussian separation, a top-hat separation, and an intermediate separation called semi-Gaussian, which is top-hat in 80\% of the redshift distribution and the rest has a smooth peak modelled by a Gaussian distribution. Our goal is to understand whether simply reducing the shot-noise is sufficient to get an accurate BAO feature or if there is a strong projection effect altering the signal.

A key question motivating this work is whether the transverse BAO can achieve percent-level accuracy for precision cosmology, or whether binning-induced 
systematics will limit its usefulness. As we will show, the choice of redshift binning can introduce projection effects that bias the recovered BAO scale by 
10--50\% or more, even in idealised scenarios. However, we also demonstrate that these biases can be controlled: for sufficiently narrow bins ($\sigma_z \lesssim 0.04$) and with the optimal binning strategy (semi-Gaussian), our proposed  correction recovers the BAO scale to percent-level accuracy. This suggests that,  with careful treatment, the transverse BAO can provide a powerful and  model-independent cosmological probe, complementary to the 3D BAO and 
particularly valuable for photometric surveys. The purpose of this work is  therefore twofold: first, to issue a quantitative warning about the dangers of 
unoptimised binning; and second, to provide a practical path forward, including a simple correction scheme and a recommendation for the semi-Gaussian bin shape.

The paper is structured as follows: in section~\ref{sec:surveys} we describe the survey configurations we will use to analyse and the fiducial model applied. Section~\ref{sec:ang} discusses how we obtain the correlation function and power spectrum, while section~\ref{sec:fisher} describes the Fisher Forecast from the simulated power spectrum. In section~\ref{sec:alpha}, we analyse the bin accuracy to find the expected $\theta_{\rm BAO}$. We check the precision of cosmological parameters in terms of the bin width in section~\ref{sec:sigma}. Finally, in section~\ref{sec:summary} we present our conclusions.

\section{Surveys and fiducial cosmology}\label{sec:surveys}

Future surveys intend to collect a large amount of observations that can be used to constrain the BAO feature. In this work, we will focus our forecast on two survey targets, the SKA HI galaxy redshift survey \cite{camera2015probing} and the DESI LRG \cite{desicollaboration2025datarelease1dark}. 

The SKA's target focusses on objects with low redshift $0<z<0.5$, while the DESI LRG is around $0.4<z<1.1$. Due to the galaxy distributions we will only use part of these redshift ranges. In terms of sky coverage, DESI intends to complete 8 million galaxies in 14,000 deg$^2$ while SKA's HI sample is limited to 5 million galaxies in 5,000 deg$^{2}$. Because we intend to analyse the implications of different bins to find the transverse BAO feature, we need a galaxy bias model for the tracers we are using, in this case, we chose SKA's model \cite{camera2015probing,bacon_ska} and LRG model by \cite{basilakos2008halo,Ntelis_2018}. A summary of the surveys specifications is found in table~\ref{tab:surveys}.

\begin{table}[ht]
    \centering
    \begin{tabular}{c|c|c}
        & SKA HI galaxy redshift survey &  DESI LRG \\
        \hline
       $N(z)$ & $10^{5.45} z^{1.31} \exp(-14.394z)$\cite{camera2015probing,bacon_ska} & Based on  DR1\cite{desicollaboration2025datarelease1dark}\\
       $b(z)$ & $a\exp(b \, z)$\cite{camera2015probing,bacon_ska} & $a\left( \frac{1+z}{1+z_c}\right)^b$\cite{basilakos2008halo,Ntelis_2018}\\
       $N_{\rm tot}$ & 5 million \cite{camera2015probing,bacon_ska} & 8 million\cite{Zhou_2023}\\
       Area (deg$^{2}$) & 5000 \cite{camera2015probing} & 14000 \cite{desicollaboration2025datarelease1dark}\\
       \end{tabular}
    \caption{Surveys' configuration.}
    \label{tab:surveys}
\end{table}

The fiducial cosmology chosen for the analysis is based on \textit{Planck}  TT, TE, EE + lowE 2018 parameters \cite{planck18}. We use the Chevallier-Polarski-Linder (CPL) model to parameterise the equation of state for Dark Energy. This model is defined by the parameters $w_0$ and $w_a$, which describe the temporal evolution of this equation of state. Specifically, the equation is expressed as $w(a)=w_0 + (1-a)\, w_a$. This approach enables a detailed analysis of how the properties of dark energy have evolved across cosmological timescales, as outlined in the works of \cite{chevallier2001accelerating,linder2003exploring}.

\begin{table}[htpb]
    \centering
    \begin{tabular}{c|c|c}
        Parameter & Fiducial choice & Differentiation step\\
        \hline
         $H_0$(km s$^{-1}$ Mpc$^{-1}$)&67.27 & 6.7  \\
         $\Omega_b h^{2}$ & 0.02236& 0.002\\
         $\Omega_c h^{2}$& 0.1202 & 0.01\\
         $w_0$& -1& 0.1\\
         $w_a$&0&0.1\\
         $a$ (SKA) & 0.616 & 0.616\\
         $b$ (SKA)& 1.017 & 1.017\\
         $a$ (DESI) & 2 & 0.2\\
         $b$ (DESI)& 1 & 0.1\\

    \end{tabular}
    \caption{{\it Left column:} Fiducial cosmological parameters adopted from {\it Planck} 2018 \cite{planck18} and our bias model. {\it Right column:} Corresponding differentiation steps for those parameters included in the Fisher analysis.}
    \label{tab:cosmo_params}
\end{table}

\section{Angular BAO feature in different bins}\label{sec:ang}

The large-scale structure of the Universe can be probed by observing the distribution of galaxies on the sky. The overdensity of galaxies in a given direction \(\hat{\mathbf{n}}\) is defined as \(\delta(\hat{\mathbf{n}})\). The two-point correlation function \(w(\theta)\), which measures the excess probability of finding two galaxies separated by an angle \(\theta\), is given by the ensemble average:
\begin{equation}\label{eq:corr_func_def}
    w(\theta) = \langle \delta(\hat{\mathbf{n}})\delta(\hat{\mathbf{n}}')\rangle,
\end{equation}
where \(\hat{\mathbf{n}} \cdot \hat{\mathbf{n}}' = \cos\theta\).

It is often convenient to work in a spherical harmonic space. The overdensity field \(\delta(\hat{\mathbf{n}})\) can be decomposed into its multipole moments \(a_{\ell m}\) via a spherical harmonic transform:
\begin{equation}\label{eq:sh_transform}
    \delta(\hat{\mathbf{n}}) = \sum_{\ell=0}^{\infty} \sum_{m=-\ell}^{\ell} a_{\ell m} Y_{\ell m}(\hat{\mathbf{n}}).
\end{equation}
The complex coefficients \(a_{\ell m}\) encode the amplitude and phase of the fluctuations. If the field is statistically isotropic, the two-point information is fully contained in the angular power spectrum \(C_\ell\), which is the variance of the \(a_{\ell m}\) for a given multipole \(\ell\):
\begin{equation}\label{eq:cl_definition}
    C_\ell = \langle a_{\ell m} a_{\ell m}^* \rangle = \frac{1}{2\ell + 1} \sum_{m=-\ell}^{\ell} |a_{\ell m}|^2.
\end{equation}

The angular power spectrum is a projection of the underlying three-dimensional matter power spectrum \(P(k)\). For a single tracer, the cross-power spectrum is given by the integral:
\begin{equation}\label{eq:cell_projection}
    C_\ell = \frac{2}{\pi} \int \mathrm{d}k\, k^2 P(k) \Delta_\ell^2(k) .
\end{equation}
Here, \(\Delta_\ell(k)\) is the transfer function or radial kernel that projects the 3D Fourier modes onto the 2D sphere. For a galaxy survey with a radial selection function \(\phi(r)\), the transfer function can be expressed in terms of a window function as:
\begin{equation}\label{eq:transfer_function}
    \Delta_\ell(k) = \sqrt{\frac{2}{\pi}} \int \mathrm{d}r\, r^2 \phi(r) j_\ell(kr) W(r),
\end{equation}
where \(j_\ell(kr)\) is the spherical Bessel function of order \(\ell\), and the window function \(W(r)\) represents any additional weighting or filter applied to the density field. This window function can be expressed in redshift ($W(z)$) instead of $r$, and its shape is what we call the bin choice to analyse whether it is or not a good function to represent a central redshift $z_c$. The radial selection function \(\phi(r)\) encodes the probability of observing a galaxy at comoving distance \(r\), normalised such that \(\int \mathrm{d}r\, r^2 \phi(r) = 1\).

\begin{figure}[ht]
    \centering
    \begin{minipage}{0.45\textwidth}
        \centering
        \includegraphics[width=\linewidth]{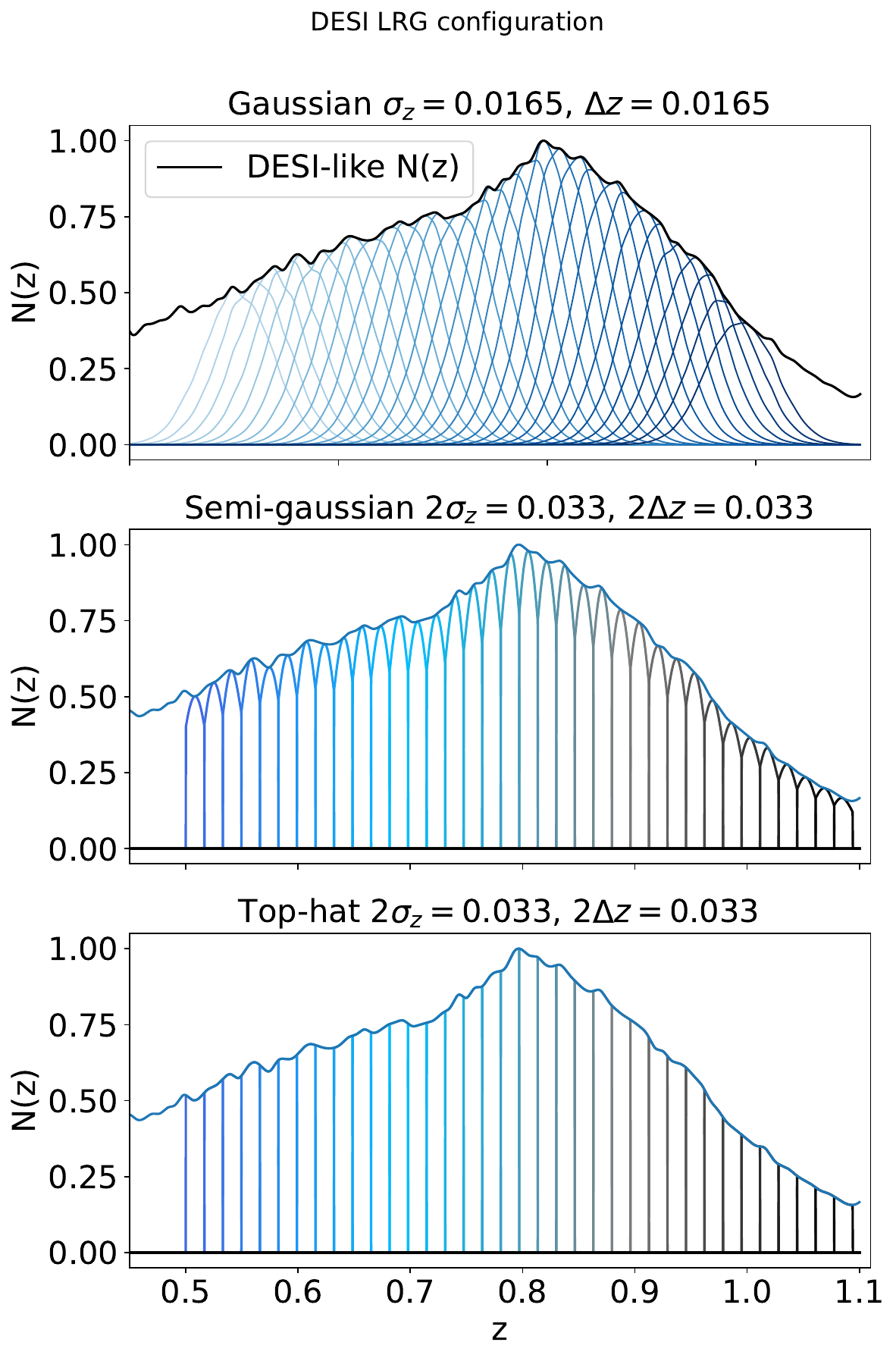}
        \caption{DESI LRG redshift distribution separated in different redshift bins, Gaussian (upper panel), Semi-Gaussian (middle panel), and Top-hat (bottom panel). The colours from blue to black represent and increase in redshift.}
        \label{fig:desi_bins}
    \end{minipage}
    \hspace{5mm}
    \begin{minipage}{0.45\textwidth}
        \centering
        \includegraphics[width=\linewidth]{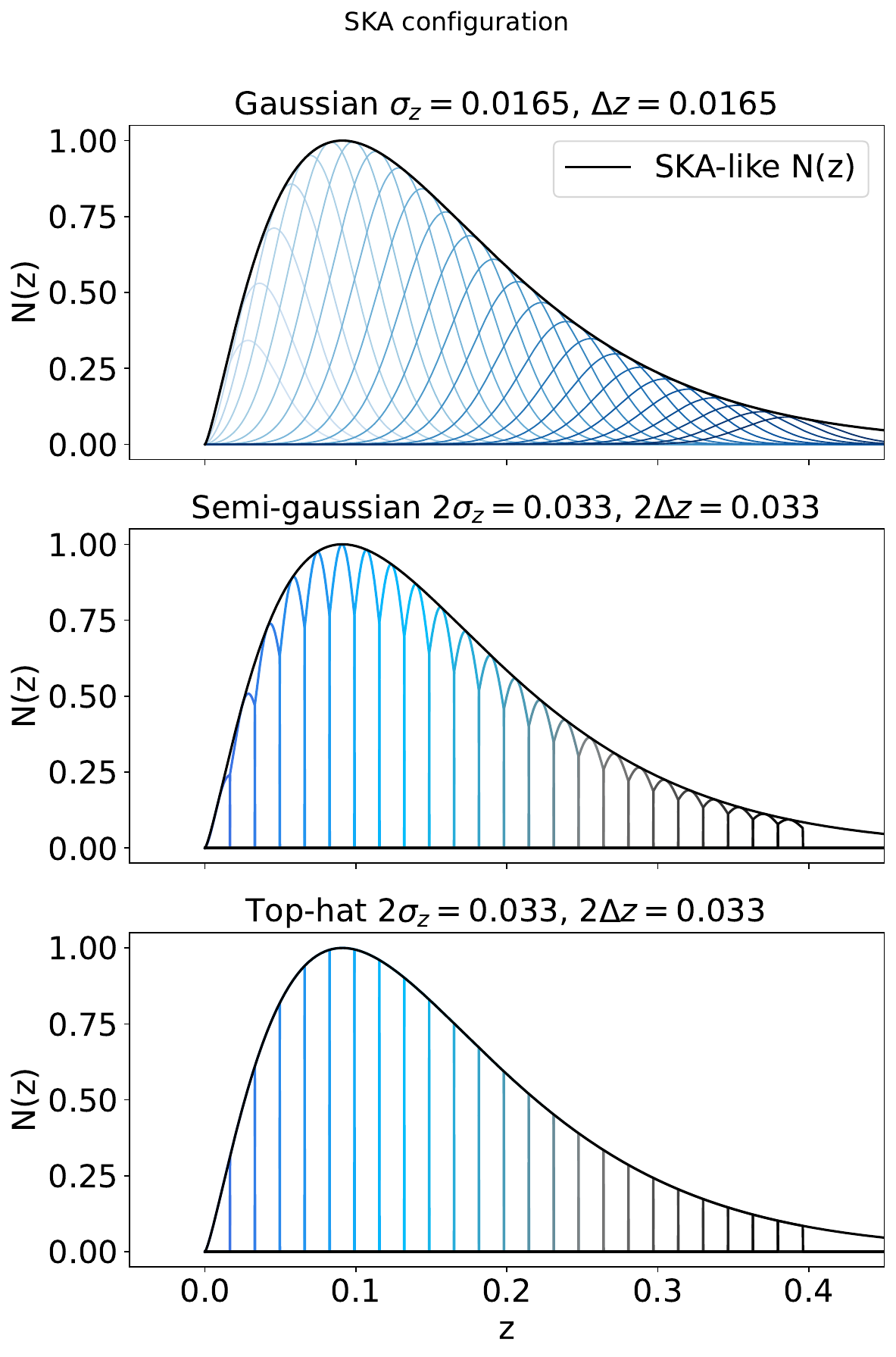}
        \caption{SKA redshift distribution separated in different redshift bins, Gaussian (upper panel), Semi-Gaussian (middle panel), and Top-hat (bottom panel). The colours from blue to black represent and increase in redshift.}
        \label{fig:ska_bins}
    \end{minipage}
\end{figure}

In this work, we model the non-linear matter power spectrum \(P(k)\) using the updated halo model prescription from \cite{Mead_2021} used in \texttt{CAMB} \cite{Lewis:1999bs}. This model modifies the linear power spectrum \(P_{\mathrm{lin}}(k)\) to account for non-linear damping of the BAO:
\begin{equation}\label{eq:hmcode_pk}
    P(k) = \left[ (1 - f_{\mathrm{nl}}) + f_{\mathrm{nl}} \left( \frac{1}{1 + (k / k_{\star})^4} \right) \right] P_{\mathrm{lin}}(k),
\end{equation}
where \(f_{\mathrm{nl}}\) and \(k_{\star}\) are simulation-calibrated parameters.

Finally, the angular correlation function \(w_i(\theta)\) of bin $i$ and the angular power spectrum \(C_\ell^i\) form a Fourier pair on the sphere. Using the addition theorem for spherical harmonics, one can relate them via a Legendre series:
\begin{equation}\label{eq:wtheta_from_cl}
    w_{i}(\theta) = \sum_{\ell=0}^{\infty} \frac{2\ell + 1}{4\pi} C_\ell^{i} P_\ell(\cos\theta),
\end{equation}
where $P_\ell$ is the Legendre polynomial of degree \(\ell\). When the angular power spectrum $C_\ell$ is computed from a very thin redshift bin, the radial projection kernel is sharply peaked in comoving distance, so that under the Limber approximation $C_\ell \propto P(k=\ell/\chi)$ directly inherits the full oscillatory BAO feature of the three‑dimensional power spectrum $P(k)$. Transforming such an oscillatory $C_\ell$ to the correlation function $w(\theta)$ involves an integral over a finite range of multipoles (e.g., $\ell_{\min}$ to $\ell_{\max}$). The abrupt truncation of the Fourier–Bessel (or Hankel) transform acts as a sharp window in $\ell$‑space, producing a ringing pattern around the BAO peak. This Gibbs phenomenon generates multiple spurious sidelobes that are easily misinterpreted as additional peaks. 

The transverse BAO feature $\theta_{\rm BAO}$ is written as 
\begin{equation}\label{eq:bao}
    \theta_{\rm BAO}=\frac{r_s}{(1+z) D_A(z)}
\end{equation}
where $r_s$ is the sound horizon at the instant baryons and photons no longer work as a fluid and $D_A$ is the angular diameter distance. It is clear that the BAO transverse scale is larger for low redshift values. In \cite{ferreira2025arf}, the authors discuss this particular dependence on the redshift in terms of the width of a particular bin, for higher $z_c$, the width must be wider in order to detect the BAO feature because the redshift is directly related to the line-of-sight direction. In a spatially flat Friedmann–Lemaître–Robertson–Walker (FLRW) universe, an observer located at redshift $\tilde{z}$ relates a comoving radial distance $r_s$ to a redshift interval $(\Delta z)^{BAO}$ via
\begin{equation}\label{eq:comoving}
    r_s = cH_0^{-1}\int_{\tilde{z}}^{\tilde{z}+(\Delta z)^{BAO}}\frac{dz'}{E(z')},
\end{equation}
where $E(z)\equiv\sqrt{\Omega_m (1+z)^3 + \Omega_{\rm DE}(z)}$. Here, $\Omega_m$ is the matter density parameter, and $\Omega_{\rm DE}(z)$ is the dark energy density parameter. For $\tilde{z}=0$, one obtains $(\Delta z)^{BAO} \simeq 0.034$ for $r_s\simeq 100~\text{Mpc}~h^{-1}$, with $(\Delta z)^{BAO}$ increasing as $\tilde{z}$ grows. If the shell width $\Delta z$ is too narrow, the shell contains too few galaxies, resulting in poor statistics and a weak correlation function signal. Conversely, if $\Delta z$ is too wide, the projection averages over a large cosmic volume, smoothing out the BAO feature due to the evolution of the angular diameter distance and the growth of structure across the bin. Setting $\Delta z$ to a value just below $(\Delta z)^{\text{BAO}}(\tilde{z})$ ensures that the shell is thick enough to contain a sufficient number of galaxies for a stable clustering measurement, yet thin enough to prevent significant cosmological evolution from smearing the characteristic BAO peak in the angular correlation function.

This dependence on width was explored in \cite{Sanchez_2010}, a clear deviation from the expected $\theta_{\rm BAO}$ appears when $z_c$ is low and the width of the bin increases. The selection of width significantly influences the resulting projection effect, directly impacting how the projection is perceived. The authors in \cite{Sanchez_2010} analysed the shift of the parametrised correlation function found whose peak is located at $\theta_{\rm fit}$ from the expected $\theta_{\rm BAO}$ through a correction scale leading to $\theta_{\rm BAO}=\alpha \theta_{\rm fit}$. However, the influence on the bin's shape was never explored in the literature, the shape not just has a significant relationship with the shot-noise of each bin but also implies changes on $z_c$'s relevance.

The possible sensitivity to the bins' width and shape is our main concern in this work. We tested three configurations of redshift bin, one is the Gaussian bin, as in Eq.~\ref{eq:gaussian_bin}, and the other two are top-hat and semi-Gaussian bins. Top-hat bins are simply a raw separation of redshift in a survey $N(z)$, whereas the semi-Gaussian is like a top-hat separation up to 80\% of the bin distribution's height, where the rest is similar to a Gaussian distribution ($n_i(z)$) peaked at $z_c$. These last two configurations can be summarised by Eq.~\ref{eq:semi_gauss}, when $A=0.2$, $B=0.8$ and $\tilde{\sigma}_z=0.01$ for semi-Gaussian and $\tilde{\sigma}_z=1$ for the top-hat bins.
\begin{equation}\label{eq:gaussian_bin}
    n_i(z)=N(z) \exp\left(\frac{-(z-z_c^i)^2}{2\sigma_z^2}\right)
\end{equation}

\begin{equation}\label{eq:semi_gauss}
    \tilde{n}_i(z)=N(z) \left[ A+B\exp\left(\frac{-(z-z_c^i)^2}{2\tilde{\sigma}_z^2}\right)\right]
\end{equation}

Figures \ref{fig:desi_bins} and \ref{fig:ska_bins} show examples of bin configurations that we will use for the analysis of a fixed bin width. We tried to keep the full width the same for all configurations, so the Gaussian $\sigma_z$ is half $\tilde{\sigma}_z$. The top-hat and semi-Gaussian have the same width in all cases. The reason for this is that it is not possible to form top-hat, Gaussian, and semi‑Gaussian binning schemes that yield the same effective number density while preserving their distinct functional forms and a common width parameter $\sigma_z$. The effective number of galaxies assigned to a bin centred at $z_c$ is given by the integral of the weighting function $W(z;z_c,\sigma_z)$ multiplied by the true galaxy density $N(z)$. Even under the simplifying assumption of a constant $N(z)=n_0$, the integrals of the three physically motivated windows differ intrinsically. Even if we restrict attention to the central region where the windows are non‑negligible, these integrals are manifestly different unless we redefine $\sigma_z$. 

\begin{figure}[ht]
    \centering
    \begin{minipage}{0.45\textwidth}
        \centering
        \includegraphics[width=\linewidth]{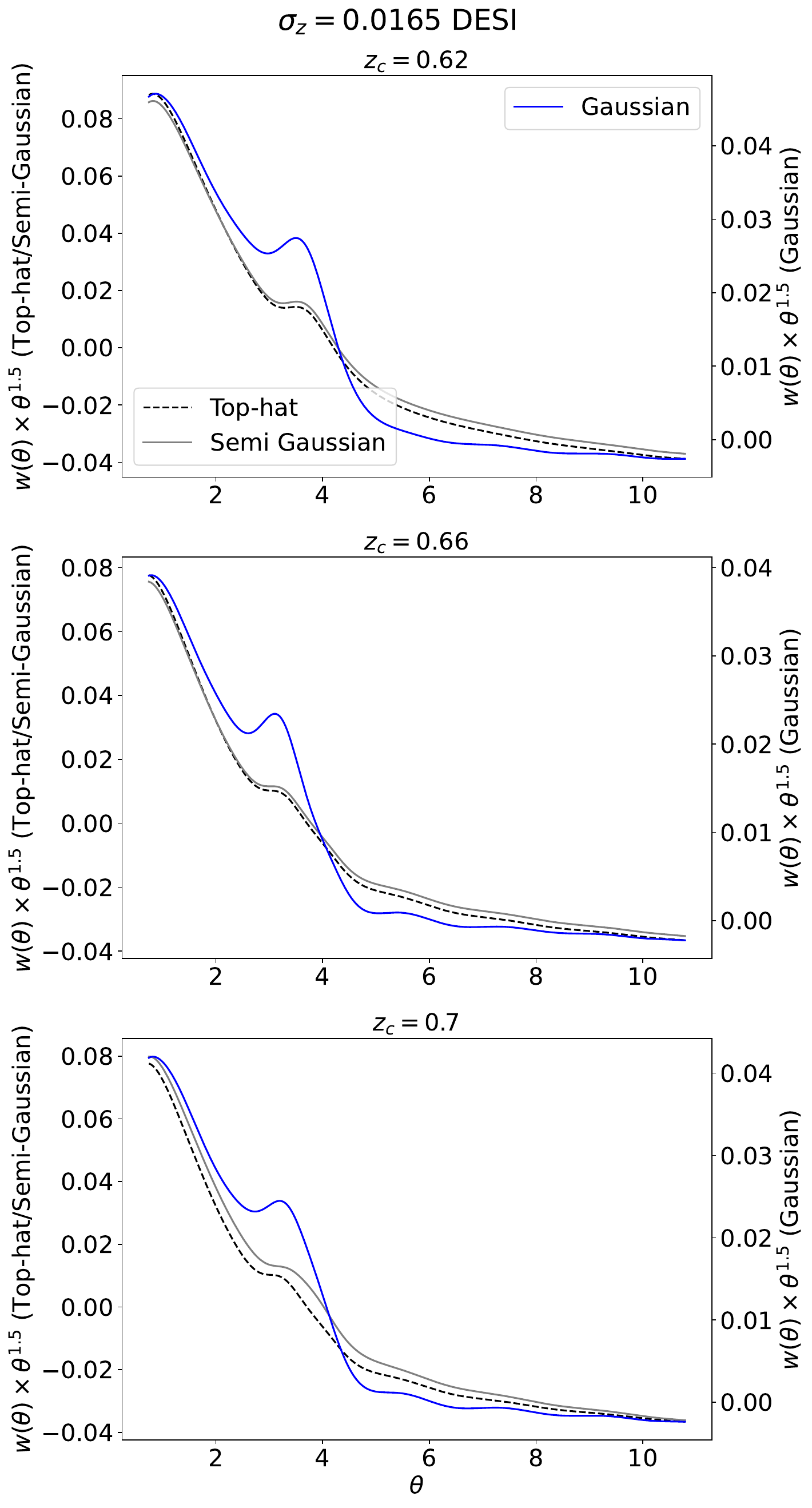}
        \caption{Angular correlation function for different bin configuration for a DESI-like survey. The Top-hat (black dashed line) and semi-Gaussian (gray line) is in the left y-axis while the Gaussian (blue line) is the right one in all panels. The upper panel has $z_c=0.62$, the middle one $z_c=0.66$, and the bottom panel has $z_c=0.7$.}
        \label{fig:desi_bins2}
    \end{minipage}
    \hspace{5mm}
    \begin{minipage}{0.45\textwidth}
        \centering
        \includegraphics[width=\linewidth]{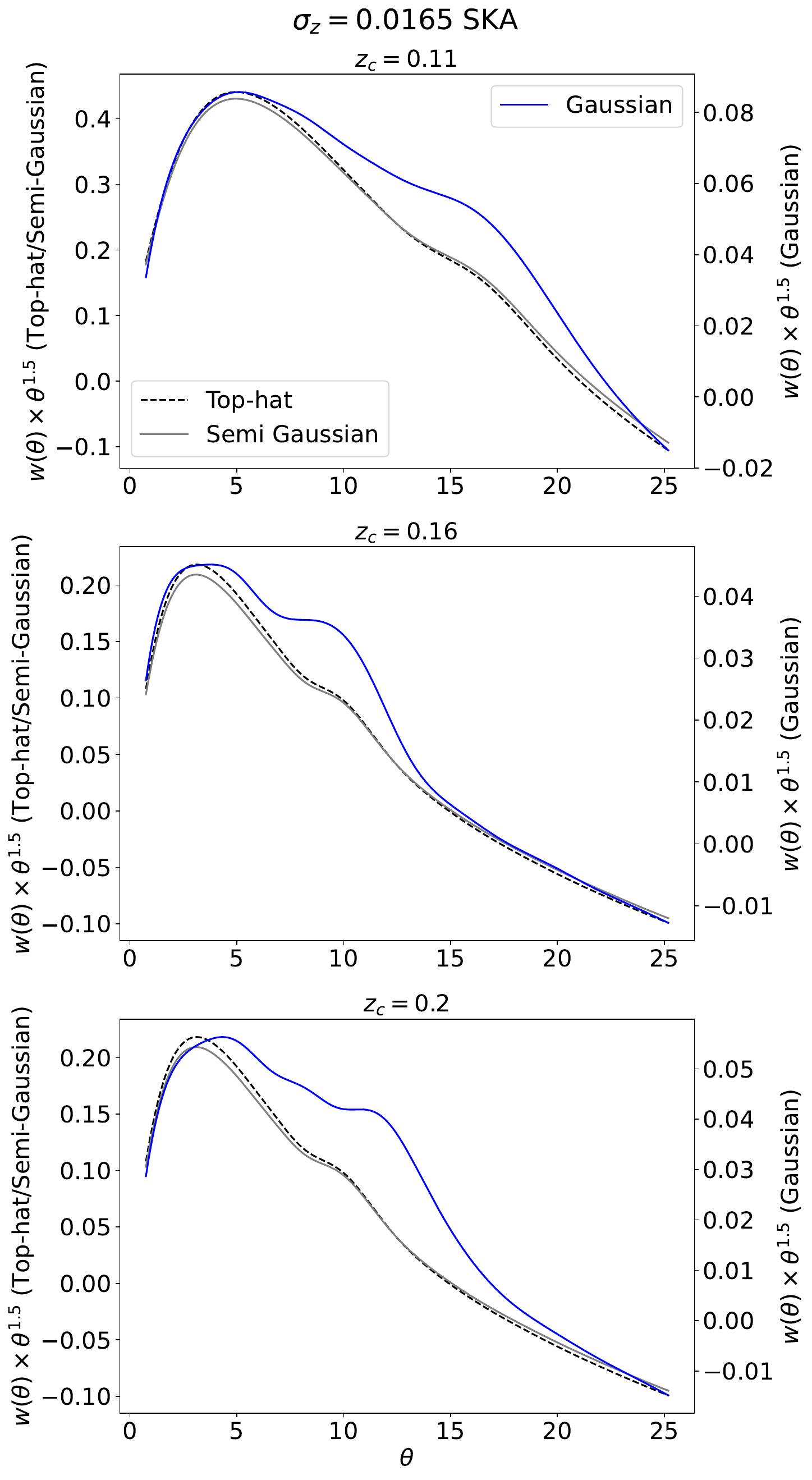}
        \caption{Angular correlation function for different bin configuration for a SKA-like survey. The Top-hat (black dashed line) and semi-Gaussian (gray line) is in the left y-axis while the Gaussian (blue line) is the right one in all panels. The upper panel has $z_c=0.11$, the middle one $z_c=0.16$, and the bottom panel has $z_c=0.2$.}
        \label{fig:ska_bins2}
    \end{minipage}
\end{figure}

The resulting $w(\theta)$ for each bin configuration can be found in Fig.~\ref{fig:desi_bins2} and \ref{fig:ska_bins2}, where we chose from a lower to a higher $z_c$. The Top-hat (black dashed line) and semi-Gaussian (gray line) is in the left y-axis while the Gaussian (blue line) is the right one in all panels with a higher amplitude. In DESI configuration, the upper panel has $z_c=0.62$, the middle one $z_c=0.66$, and the bottom panel has $z_c=0.7$. It is evident a slight displacement in the BAO peak position between the semi-Gaussian/Top-hat and the Gaussian bin schemes, this is a clear evidence of the bin's influence in the projection effect. For the SKA configuration, we follow the same colour pattern and the y-axis, but now the upper panel has $z_c=0.11$, the middle panel has $z_c=0.16$, and the bottom panel has $z_c=0.2$. This low-z example shows a stronger strength dampening of the BAO feature; this aspect will be further explored in the paper. 

\section{Fisher Forecast}\label{sec:fisher}

We perform a Fisher Forecast based on the survey configurations from table~\ref{tab:surveys} and the parameters from table~\ref{tab:cosmo_params} whose differentiation step was chosen to be 10\% of the parameters values. 

The Fisher matrix is characterised as the expectation value of the curvature associated with the logarithmic likelihood function. Essentially, it measures the extent of information encapsulated in the expected data with regard to the model parameters. This approach delineates the information derived from the covariance of the data's dependence on parameters from that arising due to the mean of the data's dependence on parameters. In a variety of cosmological studies, the dominant information typically emerges from the manner in which parameters influence the covariance, specifically the power spectra. In this context, we consider the assumption that $\ell$s are independent, which enables the construction of a Gaussian covariance matrix. 

For cosmological parameters $\alpha/\beta$, we want to know how much information can be obtained from $C_\ell$ of different bins $i/j$ following the distributions of equations~\ref{eq:gaussian_bin} and \ref{eq:semi_gauss}. The measured angular power spectra in different multipole bins ($\ell$) and potentially different observables or redshift bins (i,j). The ``mean" is the theoretical power spectrum $C_\ell^i$. Thus, the Fisher matrix $F_{\alpha \beta}$ can be written as
\begin{equation}\label{eq:fisher}
    F_{\alpha \beta} = \sum_\ell \left(\frac{\partial C_\ell^i}{\partial \alpha} \mathrm{CovM}^{-1} \frac{\partial C_\ell^j}{\partial \beta}\right).
\end{equation}
$\mathrm{CovM}$ is the inverse covariance matrix of the data (we included some examples of them in Appendix~\ref{ap:covs}). It encodes the precision and correlations of the measurements. The covariance (Eq. \ref{eq:cov_mat}) tells you how much uncertainty there is in your measurement of $C_\ell^i$ and how much it correlates with the measurement of $C_\ell^j$. 
\begin{equation}\label{eq:cov_mat}
    \mathrm{CovM}[C_{\ell}^{i,j},C_{\ell}^{i',j'}] = \frac{C_{\ell}^{i,i'}C_{\ell}^{j,j'}+C_{\ell}^{i,j'}C_{\ell}^{j,i'}}{f_{\rm sky}(2\ell+1)}.
\end{equation}
Taking the inverse ($\mathrm{CovM}^{-1}$) effectively ``weights" the Fisher matrix calculation: more precise measurements (lower variance) contribute to more information.

The density of galaxies has a direct impact on the power spectra, which needs the inclusion of the shot noise contribution. This is represented by the inverse of the galaxy number density within a particular redshift bin $i$:
\begin{equation}\label{eq:shot_noise1}
    \mathcal{N}_i=\frac{1}{N^i_g},
\end{equation}
In this equation, $N^i_g$ is calculated as $n^i/4\pi f_{sky}$, where $N_g$ is the total sum $\sum_iN^i_g$, representing the galaxy density per unit solid angle on a sphere, while $n^i$ denotes the number of galaxies present within bin $i$. As a result, it is necessary to include each $\mathcal{N}_i$ in $C_\ell^i$ in order to account for the shot-noise accurately, the usual effect is to wash-out the BAO signal, but in our analysis we found that the contribution is minimum for both surveys.

\begin{figure}[ht]
    \centering
    \includegraphics[width=\linewidth]{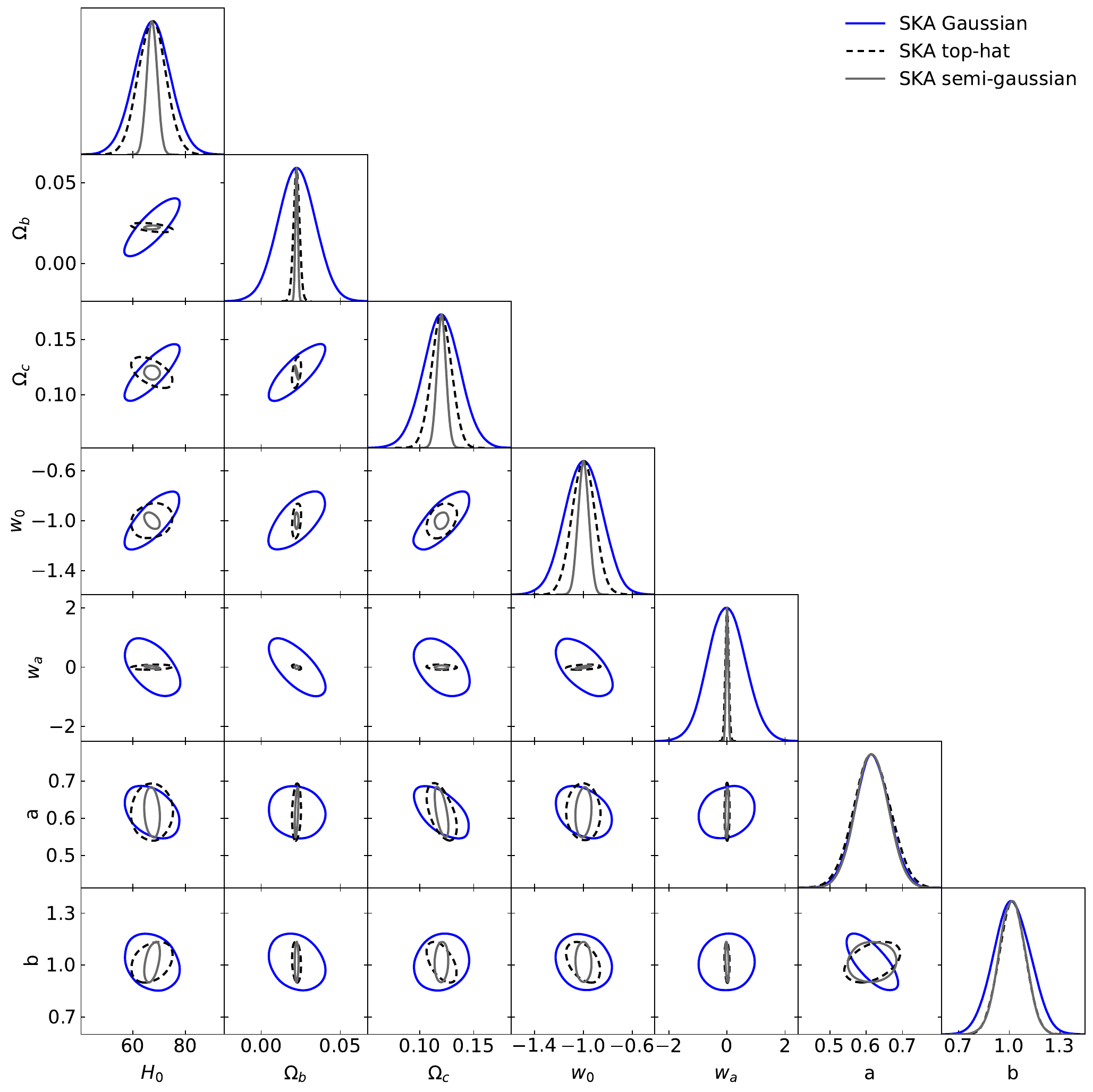}
    \caption{Triangle plot of the estimated 68\% confidence level constraints for the analysed parameters with the SKA configuration. The same colour from previous figures is followed. The covariance matrix for semi-Gaussian and top-hat schemes we multiplied by 100 to improve visualisation.}
    \label{fig:ska_forecast}
\end{figure}

From equations \ref{eq:fisher} and \ref{eq:cov_mat}, we used the SKA and DESI survey configurations for the three bins that we proposed with a similar bin width. In this case, we kept the width and bin separation equal in the Gaussian bin scheme. We conducted the forecast for a multipole range of $20<\ell<300$ where the BAO feature is more relevant to $C_\ell^i$. We also made sure to keep nearly the same $z_c$ in all bin schemes in order to provide a fair comparison. 

Figure~\ref{fig:ska_forecast} shows the resulting constraints for the cosmological parameters of interest in the SKA context. The same colour pattern from previous section is followed, for better visualisation the covariance matrix for semi-Gaussian and top-hat schemes we multiplied by 100 since the Gaussian constraint has a significant larger uncertainty. This case shows a clear better constraint of the semi-Gaussian for all parameters bin separation compared to the top-hat despite a loss in number density. 

\begin{figure}[ht]
    \centering
    \includegraphics[width=\linewidth]{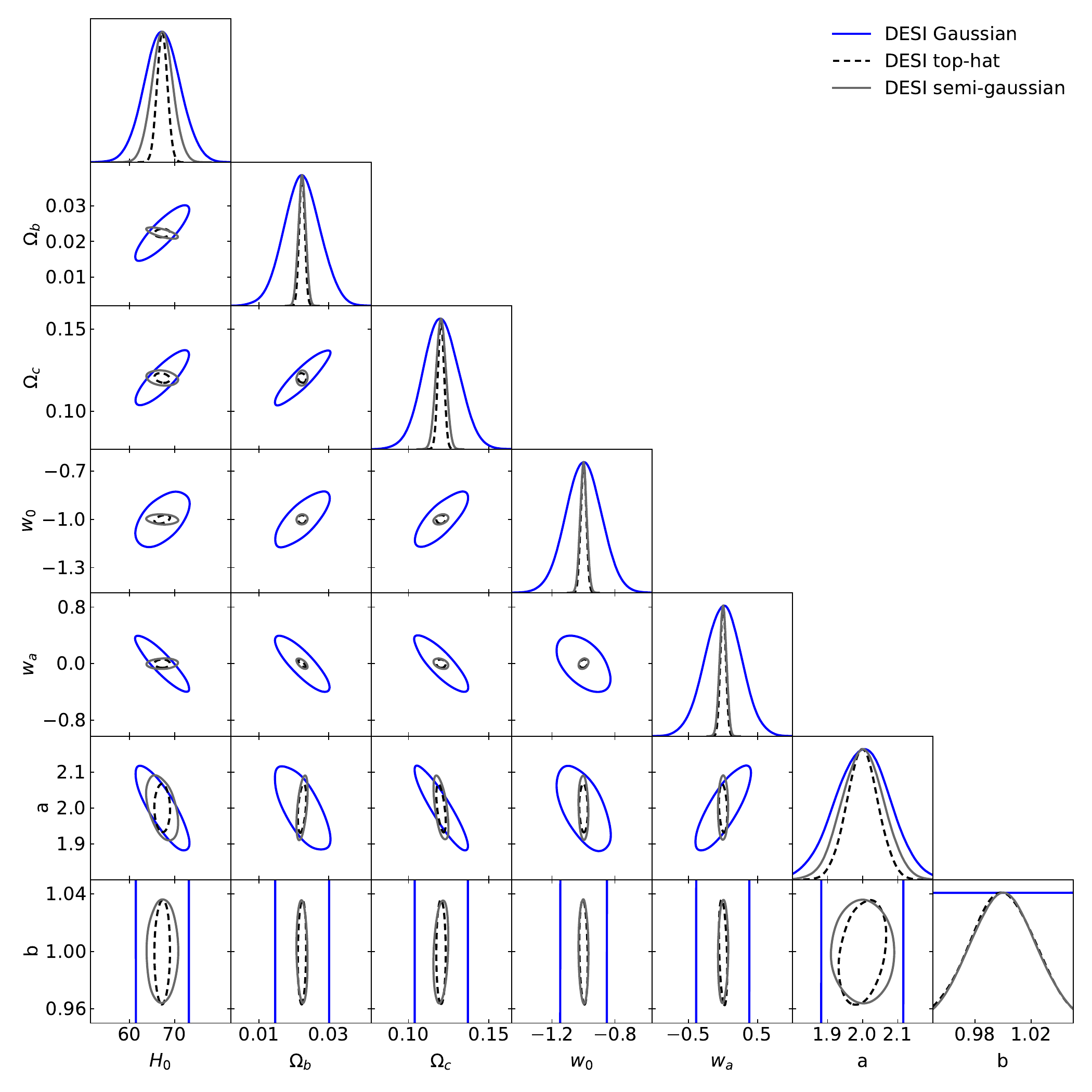}
    \caption{Triangle plot of the estimated 68\% confidence level constraints for the analysed parameters with the DESI configuration. The same colour from previous figures is followed.  The covariance matrix for semi-Gaussian and top-hat schemes we multiplied by 10 to improve visualisation.}
    \label{fig:desi_forecast}
\end{figure}

The same analysis made for DESI is shown in figure~\ref{fig:desi_forecast}, the covariance matrix for semi-Gaussian and top-hat schemes we multiplied by 10 to improve visualisation. Now, the best bin configuration is the top-hat case for all parameters, but the difference is minimal except for the $H_0$ parameter.

\begin{figure}[!ht]
    \centering
    \includegraphics[width=.8\linewidth]{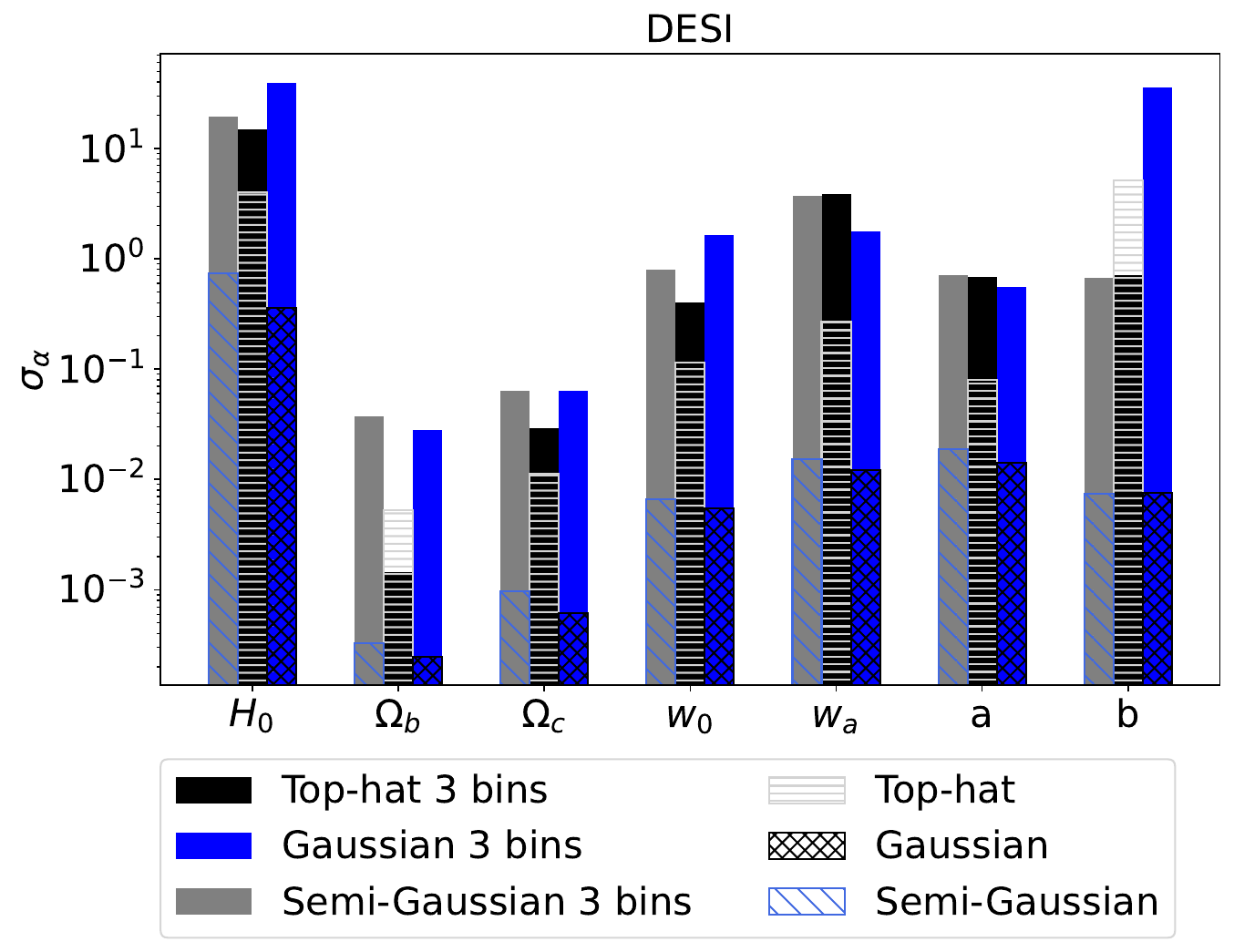}
    \caption{DESI parameters estimated errors, the solid bars represent Top-hat (in black), Gaussian (in blue), and Semi-Gaussian (in gray). The hatched patterns indicate the Fisher forecast for three bins: for the Gaussian arrangement, $z_c^g$ is set at (0.80, 0.81, 0.83), while for the top-hat and semi-Gaussian configurations, $z_c$ is adjusted to (0.80, 0.82, 0.84). Accordingly, the bars switch their colours to gray, black, and blue, in that order.}
    \label{fig:erros_desi}
\end{figure}

\begin{figure}
        \centering
        \includegraphics[width=.8\linewidth]{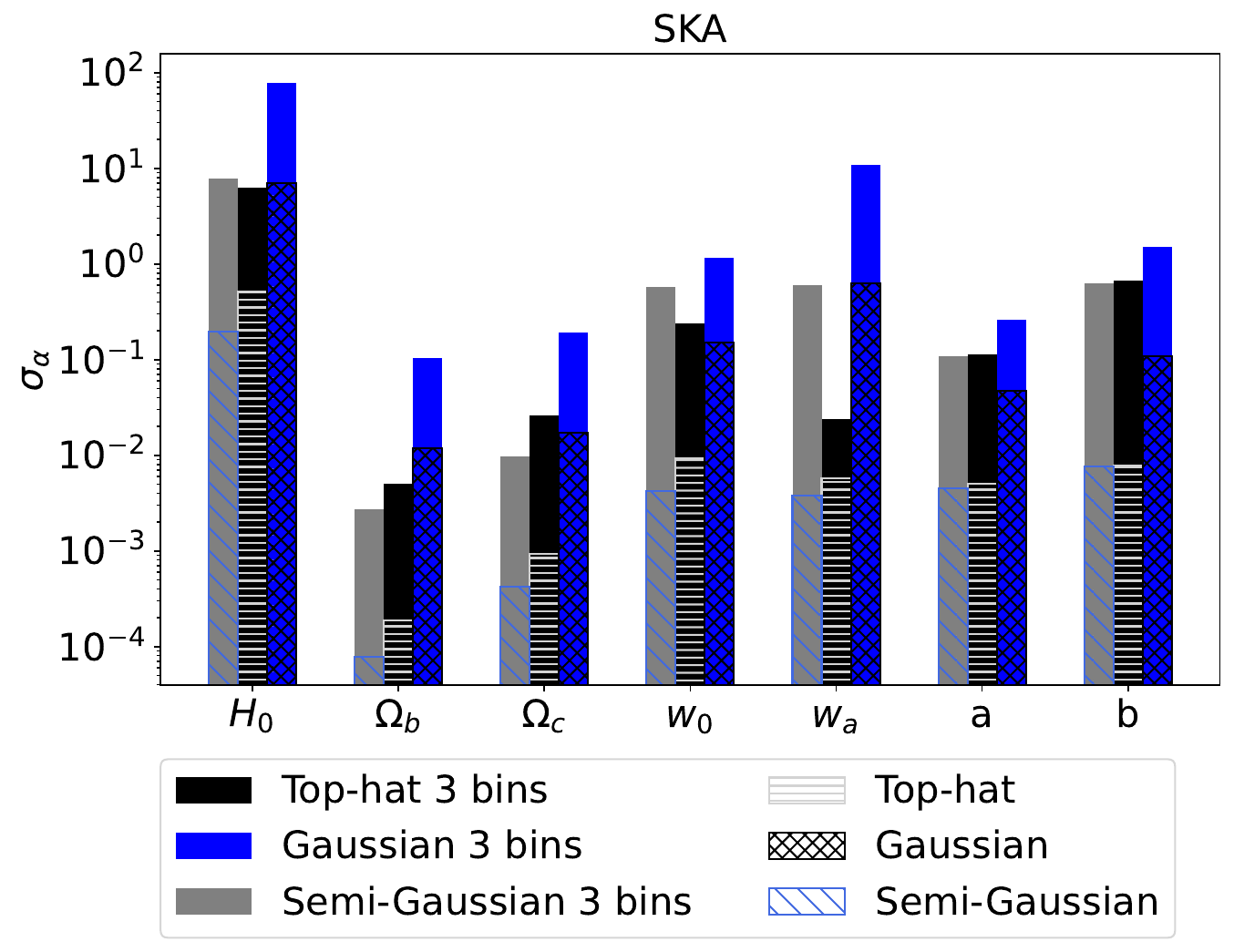}
        \caption{SKA parameters estimated errors, the solid bars represent Top-hat (in black), Gaussian (in blue), and Semi-Gaussian (in gray). The hatched patterns indicate the Fisher forecast for three bins: for the Gaussian arrangement, $z_c^g$ is set at (0.15,0.17,0.19), while for the top-hat and semi-Gaussian configurations, $z_c$ is adjusted to (0.16, 0.17, 0.19). Accordingly, the bars switch their colours to gray, black, and blue, in that order.}
        \label{fig:errors_ska}
\end{figure}

Because there is a clear difference in the bin configuration, we conducted a forecast for a smaller number of bins close to each other and located around the peak of the survey's number density. In Figures~\ref{fig:erros_desi} and \ref{fig:errors_ska}, we show the estimated errors to compare the results with a minimum number of bins, in this case three. For the DESI survey, the Gaussian bin centres are $z_c^g$=(0.80, 0.81, 0.83) and close to these values for Gaussian and Semi-Gaussian bins $z_c$=(0.80, 0.82, 0.84).  In Figure \ref{fig:errors_ska}, for the Gaussian arrangement, $z_c^g$ is set at (0.15,0.17,0.19), while for the top-hat and semi-Gaussian configurations, $z_c$ is adjusted to (0.16, 0.17, 0.19). We see that for SKA semi-Gaussian has the best constraints for all parameters in the many bins case. Moreover, as expected, the error bars grow substantially when the number of bins is reduced. While ideal tomography would employ as many bins as possible, we consider this test essential to determine whether a central, higher-density bin is disproportionately affecting the constraints.


\section{The correct BAO position}\label{sec:alpha}

The Fisher Forecast is crucial to understand which configuration improves the parameter estimation, but we lose the perception on the BAO feature detection. We use equation~\ref{eq:wtheta_from_cl} for a series of $\sigma_z$ with separation $\Delta z = 2\sigma_z$ and find the probable BAO peak, we will name $\theta_{\rm fit}$ as in \cite{Sanchez_2010}. We estimate the BAO angular scale using a hybrid, model‑free method that combines information from the angular power spectrum $C_\ell$ and the angular correlation function $w(\theta)$. After that, we compare its precision to the fiducial prediction $\theta^{\rm fid}_{\rm BAO}$ that would be obtained in the absence of projection effects (equation \ref{eq:bao}). We will use $\alpha$ as in $\theta_{\rm fit}=\alpha \theta_{\rm BAO}$ to estimate the deviation from the fiducial value.

For this analysis we estimate the BAO angular scale using a hybrid, model‑free procedure that operates simultaneously on the angular power spectrum $C_\ell$ and the pre‑weighted angular correlation function $w(\theta) \times \theta^3$. The two independent estimates are later combined by a confidence‑weighted average. We restrict the multipole range to $\ell\in[5,700]$ (the BAO wiggles are expected in this interval for the considered redshift). The smooth, “no‑wiggle” component $C_\ell^{\rm smooth}$ is obtained by fitting a univariate smoothing spline to $\ln\ell$ versus $C_\ell$, with a smoothing parameter $s=0.7$ chosen to follow the broad shape while preserving oscillations. The fractional residual
\[
R_\ell = \frac{C_\ell - C_\ell^{\rm smooth}}{\langle C_\ell^{\rm smooth}\rangle}
\]
isolates the oscillatory part. $R_\ell$ is oscillatory with a characteristic period in $\ln\ell$ that corresponds to the BAO scale. We isolate this oscillation by applying a narrow band‑pass filter centred on the dominant frequency of $R_\ell$ (the frequency that carries most of the oscillatory power). The result is a filtered residual $\widetilde{R}_\ell$ that contains only the BAO wiggles. The first positive maximum of $\widetilde{R}_\ell$ is located at multipole $\ell_{\rm BAO}$, and the angular scale is obtained as $\theta_{\rm BAO}^{(C)} = 180^\circ / \ell_{\rm BAO}$. The confidence $c_C$ of this measurement is the height of that peak divided by the standard deviation of $\widetilde{R}_\ell$.

The input correlation function has been multiplied by $\theta^3$ to enhance the BAO bump. Over the angular range $\theta\in[0.5^\circ,20^\circ]$ we apply a light Gaussian smoothing ($\sigma = 0.5$ bins) to reduce numerical noise. We then locate the first positive peak of the smoothed $\theta^3 w(\theta)$; this peak is identified as $\theta_{\rm BAO}^{(w)}$. Its confidence $c_w$ is the prominence of the peak divided by the standard deviation of the smoothed curve. If no positive peak is found, we fall back to the global maximum of the smoothed signal (provided it is positive).

The two estimates are merged as follows. If only one method succeeds, its value is returned. If both succeed, we compute the relative difference $\delta = 2|\theta_{\rm BAO}^{(C)}-\theta_{\rm BAO}^{(w)}|/(\theta_{\rm BAO}^{(C)}+\theta_{\rm BAO}^{(w)})$. When $\delta$ is below a tolerance $\epsilon=0.01$, the two measurements are considered consistent and we calculate a confidence‑weighted average:
\[
\theta_{\rm fit} = \frac{c_C\,\theta_{\rm BAO}^{(C)} + c_w\,\theta_{\rm BAO}^{(w)}}{c_C + c_w}.
\]
If $\delta \ge \epsilon$, we retain the estimate with the higher confidence. This hybrid method yields a robust, model‑free BAO angular scale that shows minimal dependence on the exact choice of the $\ell$ or $\theta$ ranges.

\begin{figure}[ht]
    \centering
    \includegraphics[width=\linewidth]{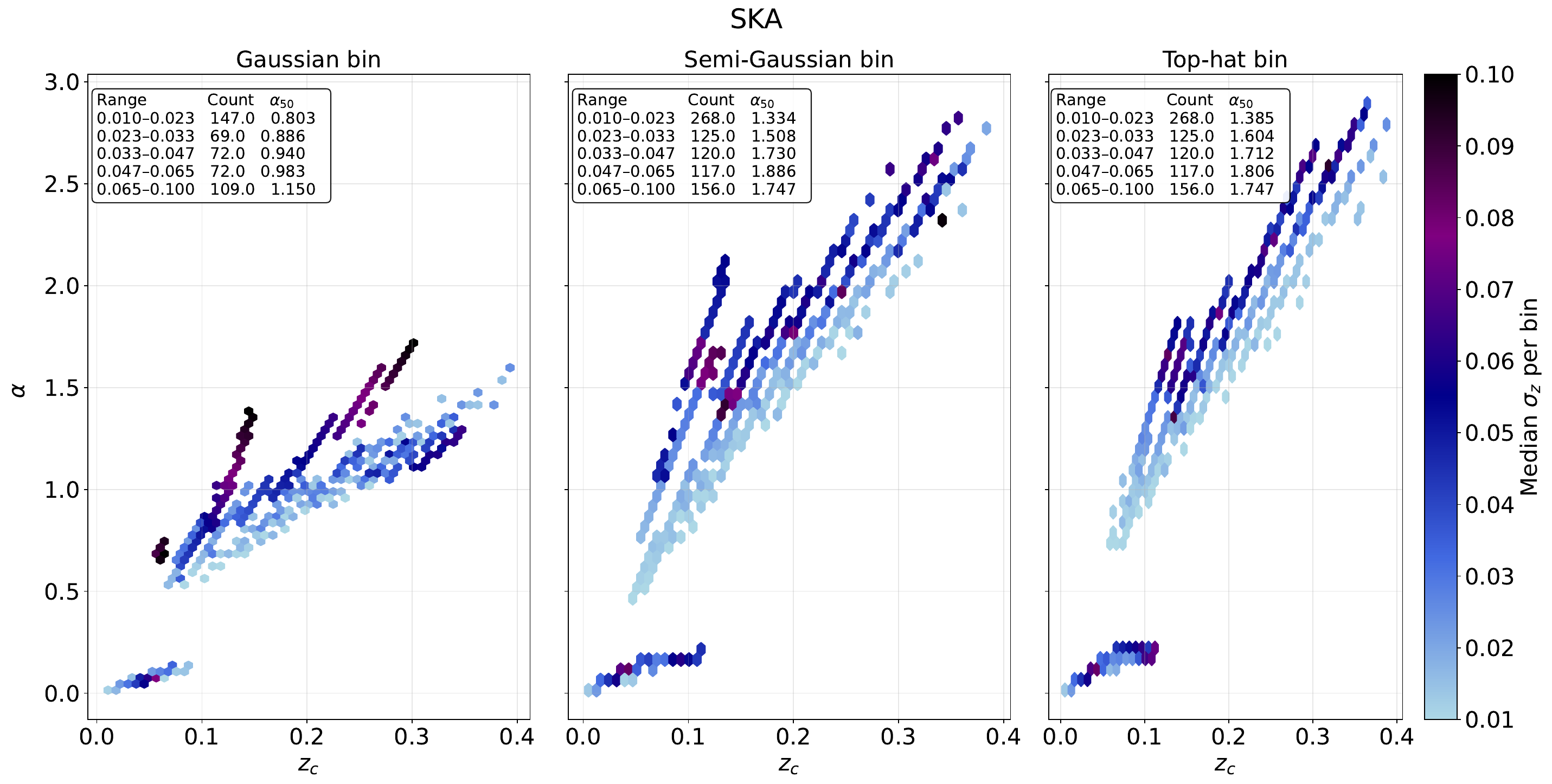}
    \caption{SKA: 2D density maps of $\alpha$ vs.\ $z_c$ for three bin‑shape assumptions (Gaussian, semi‑Gaussian, top‑hat). Colour encodes median $\sigma_z$ per bin from light blue to black. In the small panels, we display the percentile 50 of $\alpha$'s for different $\sigma_z$ ranges. Examples of individual results can be found in Appendix~\ref{ap:b}.}
    \label{fig:alpha_ska}
\end{figure}

Figure~\ref{fig:alpha_ska} shows a 2D density plot of the BAO angular scale $\alpha$ versus the bin centre $z_c$ for the SKA analysis, with the colour encoding $\sigma_z$ within each hexagonal bin. The overlaid text reports, for several $\sigma_z$ intervals (e.g., $0.010$–$0.023$, $0.023$–$0.033$, etc.), the number of points and the median $\alpha$ value. It is split into three panels corresponding to the three bin‑shape assumptions employed: Gaussian, semi‑Gaussian, and top‑hat bins. For the Gaussian bin shape, the measured $\alpha$ remains increasing but not larger than 50\% of the fiducial model. In contrast, both the semi‑Gaussian and top‑hat bins show a stronger increasing trend. This behaviour is intuitively plausible,  broader redshift scatter washes out the BAO feature, shifting the detected peak to larger angular scales. Among these two, the semi‑Gaussian bin consistently yields a smaller overestimation (i.e., $\alpha$ remains closer to the fiducial value) compared to the top‑hat bin, which exhibits a stronger positive bias at high $\sigma_z$. 

\begin{figure}[ht]
    \centering
    \includegraphics[width=\linewidth]{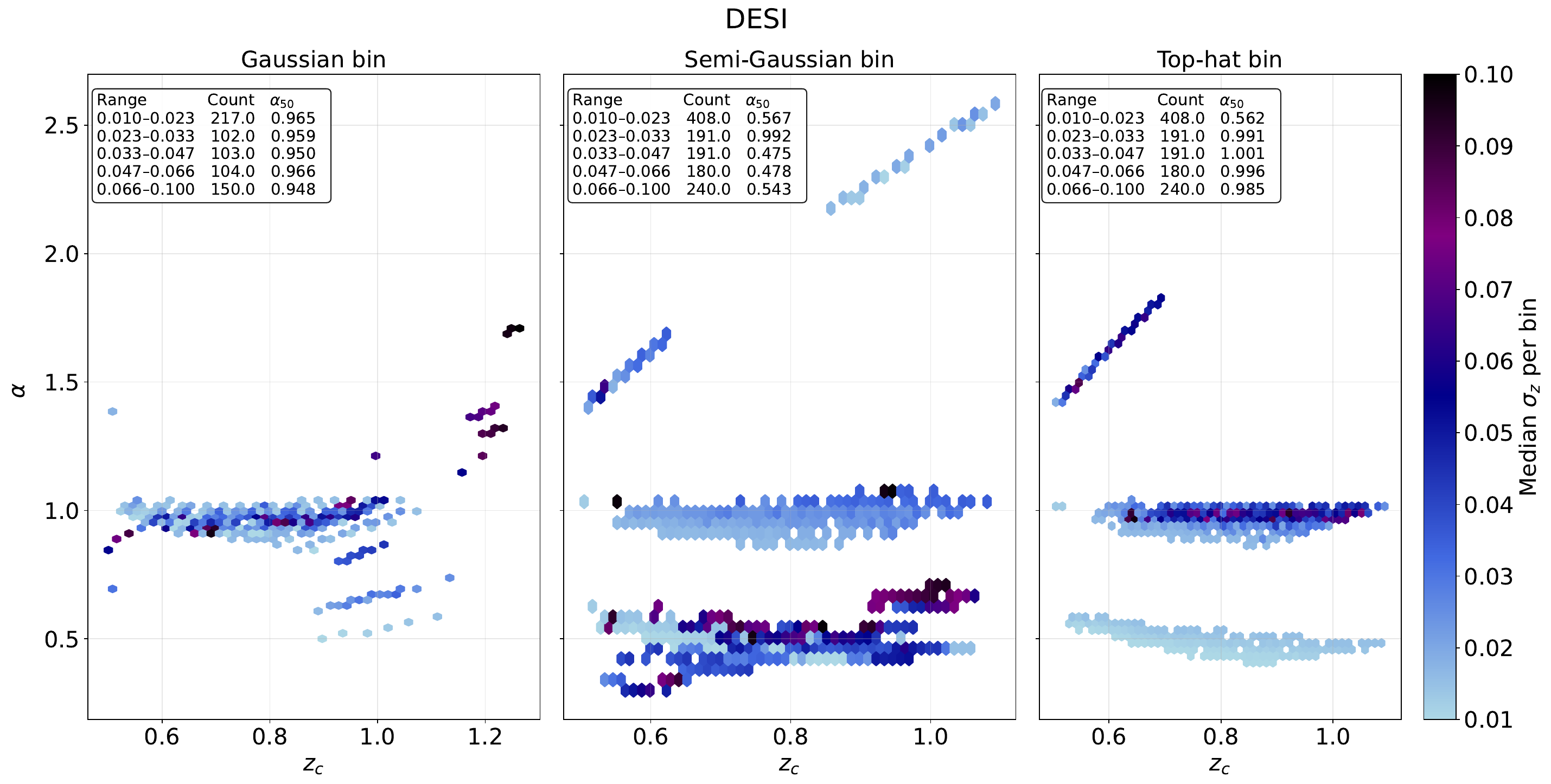}
    \caption{DESI: 2D density maps of $\alpha$ vs.\ $z_c$ for three bin‑shape assumptions (Gaussian, semi‑Gaussian, top‑hat). Colour encodes median $\sigma_z$ per bin from light blue to black. In the small panels, we display the percentile 50 of $\alpha$'s for different $\sigma_z$ ranges. Examples of individual results can be found in Appendix~\ref{ap:b}.}
    \label{fig:alpha_desi}
\end{figure}

Figure~\ref{fig:alpha_desi} presents the recovered BAO angular scale $\alpha$ as a function of $\sigma_z$ for DESI. The behaviour differs clearly between bin shapes. For the top‑hat bin, small $\sigma_z$ values lead to an underestimation of the BAO scale (i.e., $\alpha$ lower than the fiducial value). As $\sigma_z$ increases, the measurement improves, with the best result occurring in the interval $[0.033,0.047]$ where $\alpha=0.950$. This suggests that a certain amount of redshift smoothing is necessary to correctly recover the BAO feature when using a sharp bin edge. For the semi‑Gaussian bin, the opposite trend is observed: high $\sigma_z$ values cause an underestimation of the BAO scale, while lower $\sigma_z$ yields results closer to the expected value. The optimal $\sigma_z$ range for this bin shape is $[0.023,0.033]$ ($\alpha=0.959$). 
For the Gaussian bin, the behaviour is intermediate, with $\alpha$ remaining stable across all $\sigma_z$ intervals ($0.95$–$0.97$). Importantly, no universal monotonic trend exists across all bin shapes; instead, the optimal $\sigma_z$ depends on the assumed bin shape. This highlights the importance of selecting an appropriate bin shape and redshift scatter for BAO analysis.

The pronounced instability observed in the SKA survey–like outcome underscores how strongly low‑z samples are affected by the projection effect. This demonstrates that the strategy of merely collecting low‑redshift observations and then imposing an essentially arbitrary tomographic binning scheme, while presuming that it will automatically deliver a robust BAO detection, constitutes a serious oversimplification. Instead, it is crucial to carry out a systematic and quantitative optimisation of both the radial binning (i.e. the bin geometry) and the modelling of the redshift uncertainties, ensuring that the bin shape and the redshift scatter are calibrated with sufficient precision to safeguard the reliability and interpretability of the resulting BAO measurements.
For completeness, we present in Appendix~\ref{ap:b} the full set of recovered BAO angular scale $\alpha$ as a function of bin centre $z_c$ for many bin shapes and surveys. 

In \cite{Ferreira1}, the projection effect was corrected by the deviation of thin bins inside the desired bin configuration, based on the statistics of these deviations. Here, we propose a correction to the projection effect based on the fiducial BAO angular size of the adjacent parts of the bin. 

\begin{figure}
        \centering
        \includegraphics[width=\linewidth]{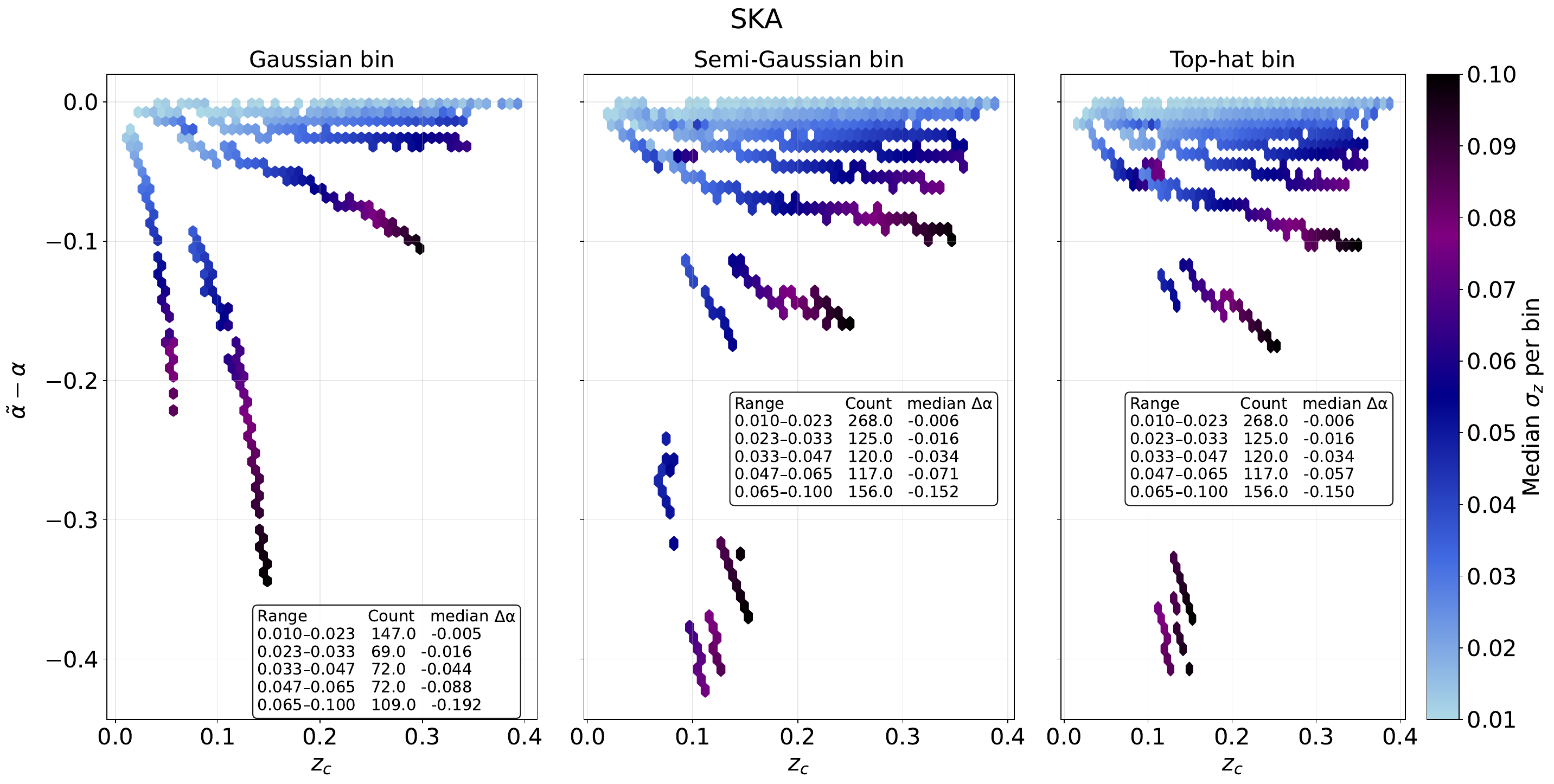}
        \caption{The discrepancy between $\tilde{\alpha}$ and $\alpha$ is illustrated for various $\sigma_z$ values, ranging from light blue to black. The left panel displays the Gaussian binning, the middle panel presents the semi-Gaussian method, and the right panel features the top-hat technique. The colours indicate the median discrepancy for different values of $\sigma_z$. As anticipated, larger $\sigma_z$ are more difficult to correct. Examples of individual results can be found in Appendix~\ref{ap:b}.}
        \label{fig:alphac1}
    \end{figure}

\begin{figure}
        \centering
        \includegraphics[width=\linewidth]{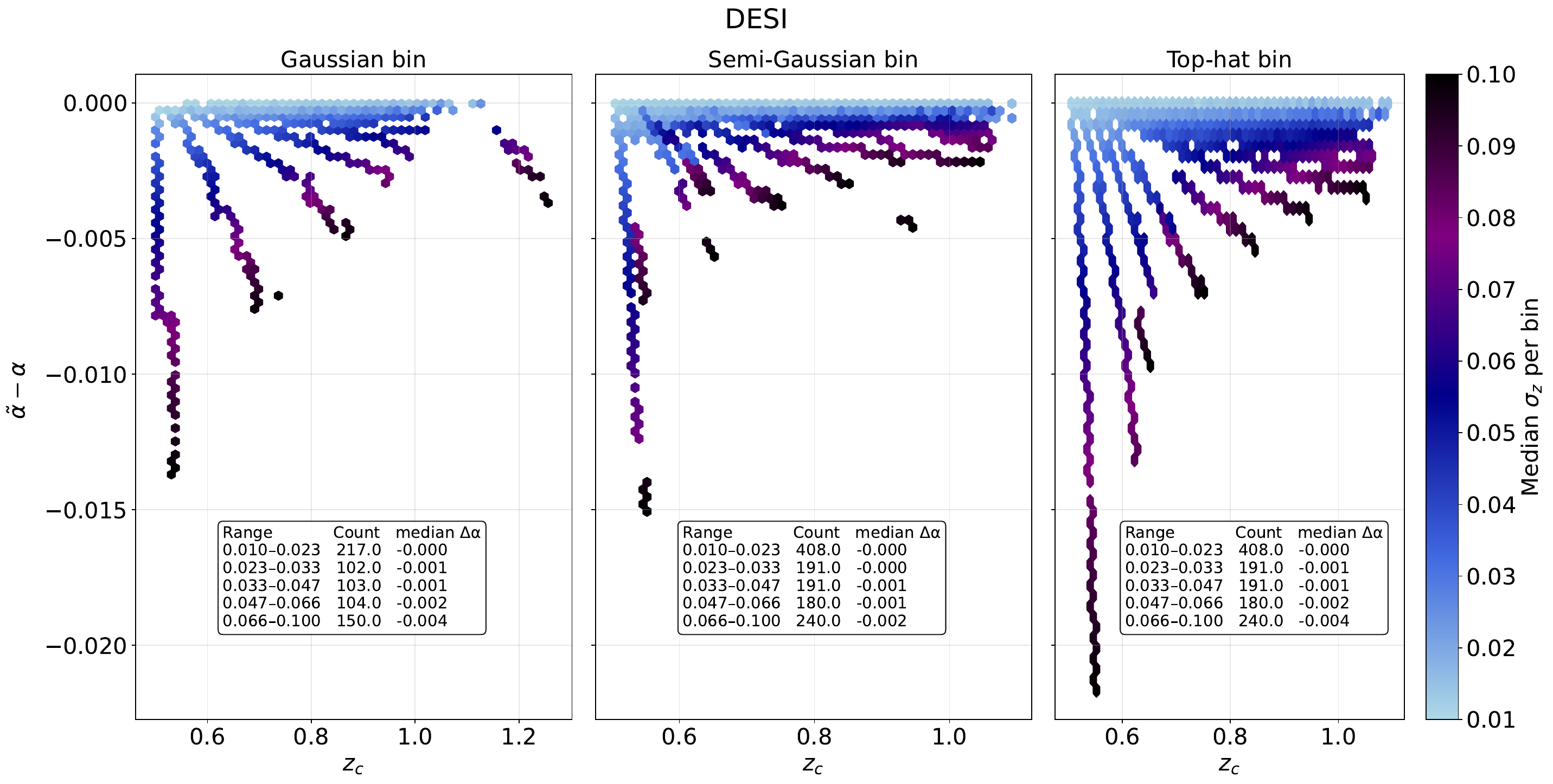}
        \caption{The discrepancy between $\tilde{\alpha}$ and $\alpha$ is illustrated for various $\sigma_z$ values, ranging from light blue to black. The left panel displays the Gaussian binning, the middle panel presents the semi-Gaussian method, and the right panel features the top-hat technique. The colours indicate the median discrepancy for different values of $\sigma_z$. As anticipated, larger $\sigma_z$ are more difficult to correct. Examples of individual results can be found in Appendix~\ref{ap:b}.}
        \label{fig:alphac2}
    \end{figure}

To mitigate the projection effect at a given central redshift $z_c$, we adopt a differential binning technique that isolates the signal intrinsic to $z_c$ by removing contamination from adjacent redshifts. 
This is achieved by considering the two halves of the redshift bin centred on $z_c$: the higher-redshift half, denoted $z_c^+$, and the lower-redshift half, denoted $z_c^-$. 
Under the assumption that the projection effect is symmetric, i.e. that the contamination entering the central bin from $z_c^+$ is comparable to the contamination leaving it toward $z_c^-$, the net contribution from neighbouring redshifts can be removed by subtracting the signal measured in the lower half-bin and that measured in the higher half-bin. Thus, the $\tilde{\theta}_{\rm BAO}$ from a projection effect correction is
\begin{equation}
    \tilde{\theta}_{\rm BAO}= \tilde{\alpha} \theta_{\rm fit}=\left[1-\alpha_{-}-\alpha_+\right]\theta_{\rm fit},
\end{equation}
where $\alpha_-$ is the contribution of the lower redshift half of the bin and $\alpha_+$ is the contribution of the higher half.
\begin{equation}
    \alpha_+=\frac{\theta^{\rm fid}_+-\theta^{\rm fid}}{\theta^{\rm fid}}
\end{equation}
\begin{equation}
    \alpha_-=\frac{\theta^{\rm fid}_--\theta^{\rm fid}}{\theta^{\rm fid}}
\end{equation}
The idea is that we can recover the $\alpha$ shift by eliminating the influence of the projection effect, thus $\alpha\sim \tilde{\alpha}$ and $\tilde{\theta}_{\rm BAO} \sim \theta_{\rm BAO}$. We emphasise that this correction does not reduce shot noise; rather, it provides a way to statistically remove the projection bias at the cost of increased noise. In practice, this trade-off may be acceptable when systematic offsets dominate over statistical uncertainties, particularly for low-redshift samples where projection effects are most severe. We compare the output of some examples in the SKA and DESI surveys through $\tilde{\alpha}-\alpha$ in Figure~\ref{fig:alphac1} and \ref{fig:alphac2}. The same colour scheme and description of the previous figures are used. In all cases, the smaller value of $\sigma_z$ more closely matches the true correction of the BAO feature. For larger $z_c$, the black hexagrams indicate that increasing $\sigma_z$ leads to an improvement in the results, as described in section \ref{sec:ang}. Nevertheless, $\sigma_z>0.05$ should not be regarded as the optimal bin width, because the corresponding correction is less efficient. We present in Appendix~\ref{ap:b} the examples of recovered BAO angular scale projection corrections $\tilde{\alpha} - \alpha$. 

To quantify the reliability of the recovered BAO scale in the presence of projection effects, we define a detection as a ``failure'' if the correction changes the angular scale by more than a given threshold. Specifically, we compute the absolute difference $|\Delta| = |\alpha - \tilde{\alpha}|$ for each case, where $\alpha$ is the uncorrected BAO scale and $\tilde{\alpha}$ is the scale obtained after applying our adjacent-bin correction (Eqs.~5.1--5.2). We then adopt an adaptive threshold $|\Delta| > 1.05 \, \langle |\Delta| \rangle$, where the mean is taken over all bin shapes and $\sigma_z$ values combined for each survey. This threshold corresponds approximately to the 95th percentile of the correction distribution and flags outliers that are more than 5\% above the typical correction.

The resulting failure rates are summarised in Table~\ref{tab:failure_rates}. For both DESI and SKA, the failure rate is negligible for narrow bins 
($\sigma_z < 0.04$), with most rates below 10\%. However, for broad bins ($\sigma_z > 0.04$), the failure rate jumps dramatically to 54--77\%, indicating that the projection correction becomes large and the BAO measurement is unreliable. This sharp increase is particularly severe for top-hat bins in the DESI survey ($76.8\%$) and for Gaussian bins in the SKA survey ($66.2\%$). In contrast, semi-Gaussian bins show the lowest failure rates for SKA in the broad-bin regime ($54.1\%$), confirming their robustness against projection effects. These results highlight that while narrow bins (regardless of shape) yield reliable BAO measurements, wide bins can produce severely biased results, and the semi-Gaussian binning offers the most stable performance when wider bins are necessary.

\begin{table}[htbp]
\centering
\caption{Fraction of BAO detections flagged as failures based on the projection correction. A failure is defined as $|\alpha - \tilde{\alpha}| > 1.05 \, \langle |\alpha - \tilde{\alpha}| \rangle$, where the mean is computed globally across all bin shapes and surveys.}
\label{tab:failure_rates}
\begin{tabular}{l l c c c}
\toprule
Survey & Bin shape & $\sigma_z < 0.02$ & $0.02 < \sigma_z < 0.04$ & $\sigma_z > 0.04$ \\
\midrule
\multirow{3}{*}{DESI} 
& Gaussian      &  0.0\% &  1.1\% & 76.6\% \\
& Semi-Gaussian &  0.0\% &  1.1\% & 76.2\% \\
& Top-hat       &  0.0\% &  1.7\% & 76.8\% \\
\cmidrule{1-5}
\multirow{3}{*}{SKA} 
& Gaussian      &  0.0\% &  3.8\% & 66.2\% \\
& Semi-Gaussian &  4.7\% &  8.7\% & 54.1\% \\
& Top-hat       &  4.7\% &  9.1\% & 55.9\% \\
\bottomrule
\end{tabular}
\end{table}

\section{Dependence on \texorpdfstring{$\sigma_z$}{sigmaz}}\label{sec:sigma}

In order to find the best $\sigma_z$, we computed the forecast for $0.01<\sigma_z<0.1$ and compared whose separation had the most information from the parameters of interest. The metric for such evaluation is the Figure of Merit (FoM) which is defined as $FoM=\sqrt{\rm det \, (F_{\alpha \beta}})$ without repeated number of bins. We tested all possible parameter pair but we will focus on: $H_0 \times w_0$, $w_0 \times w_a$, $b\times a$, and $\Omega_b \times \Omega_c$. The other parameters followed a similar pattern of FoM in terms of bin width.

\begin{figure}
    \centering
    \begin{minipage}{0.45\textwidth}
        \centering
        \includegraphics[width=\linewidth]{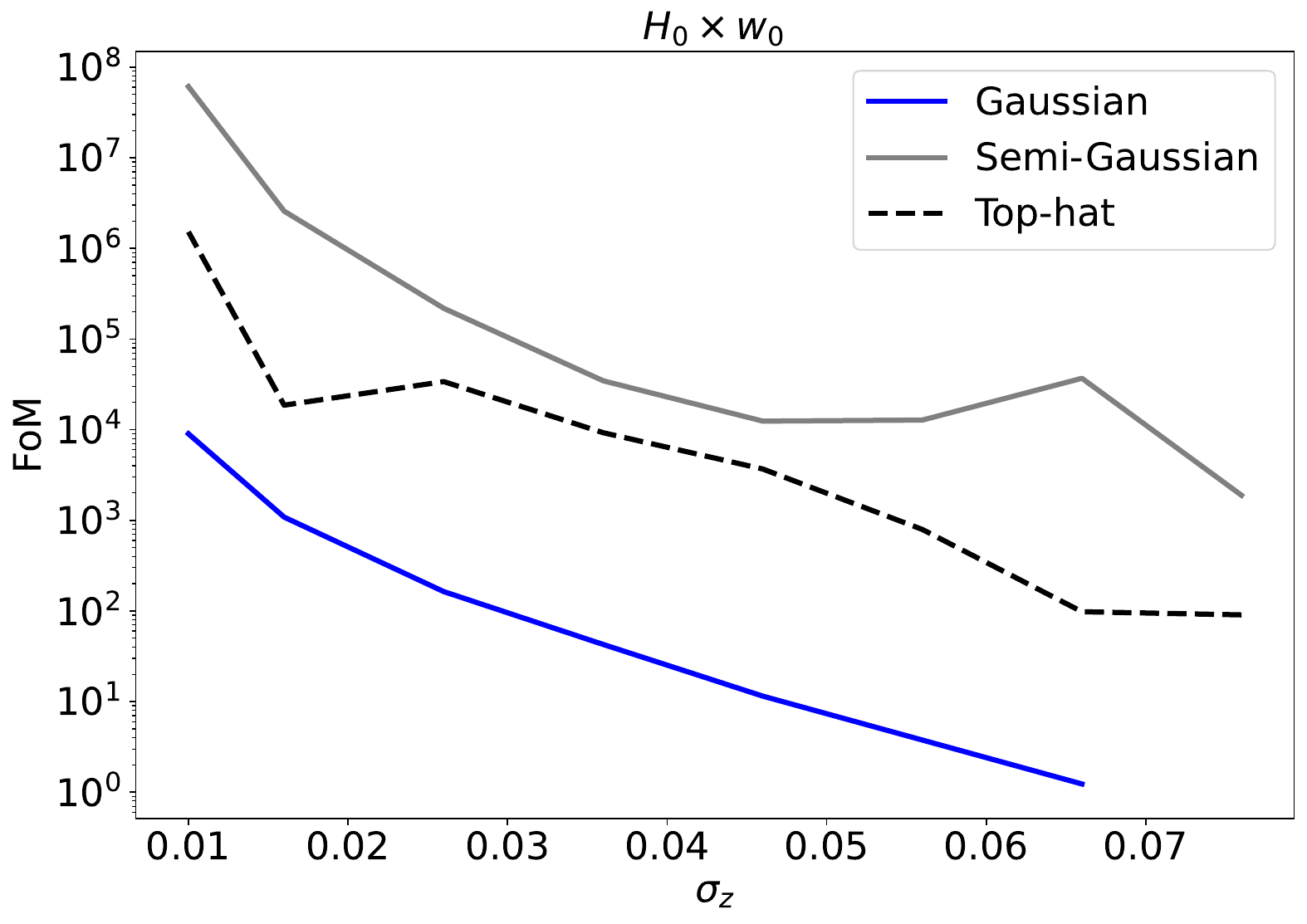}
    \end{minipage}
    \begin{minipage}{0.45\textwidth}
        \centering
        \includegraphics[width=\linewidth]{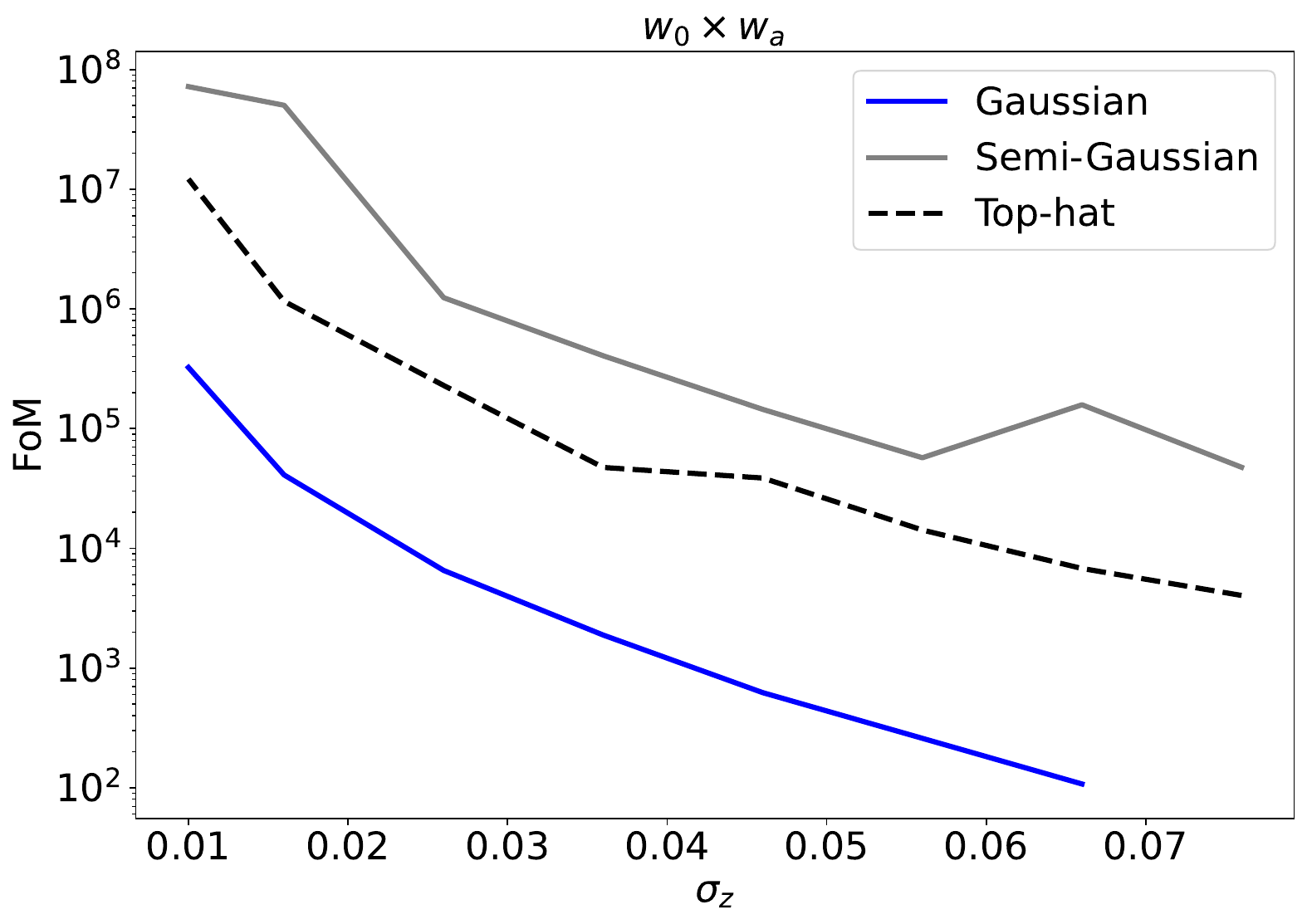}    
    \end{minipage}
    \caption{SKA FoM for different $\sigma_z$ ($\Delta z = 2\sigma_z$) for parameters $H_0\times w_0$ (left panel) and $w_0 \times w_a$ (right panel).}\label{fig:ska_fom}
\end{figure}

\begin{figure}
    \centering
    \begin{minipage}{0.45\textwidth}
        \centering
        \includegraphics[width=\linewidth]{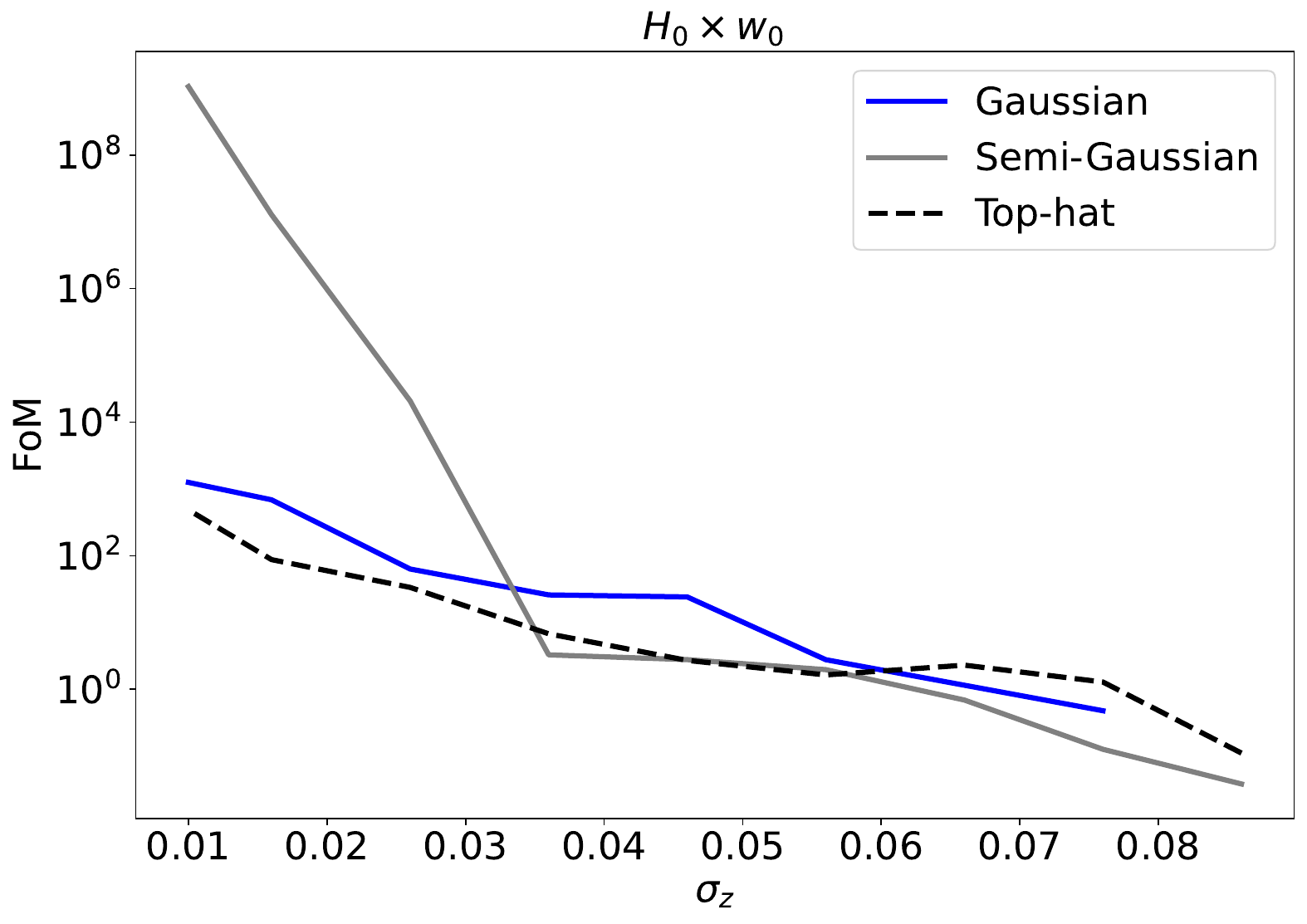}
    \end{minipage}
    \begin{minipage}{0.45\textwidth}
        \centering
        \includegraphics[width=\linewidth]{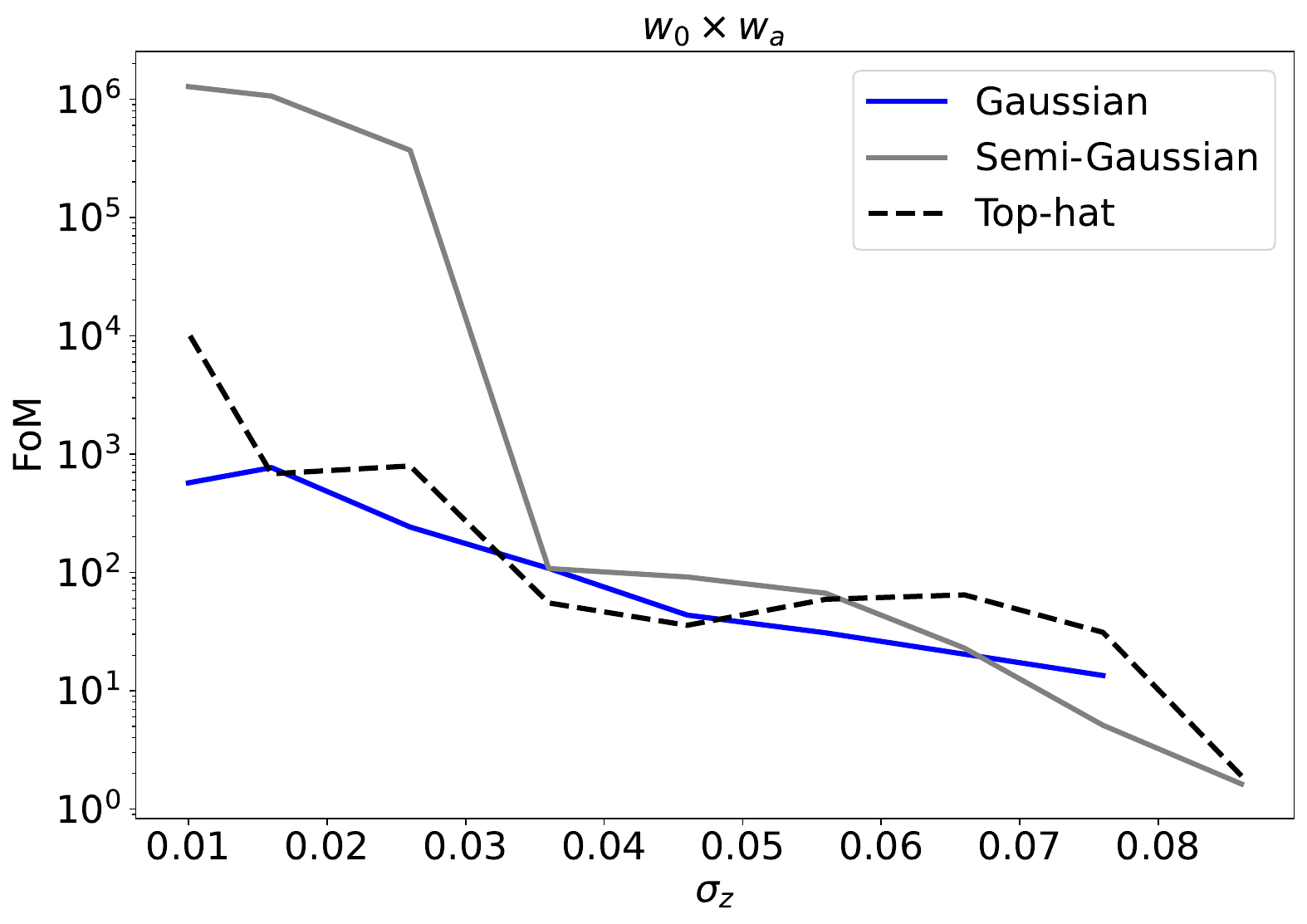}
    \end{minipage}
    \caption{DESI FoM for different $\sigma_z$ ($\Delta z = 2\sigma_z$) for parameters $H_0\times w_0$ (left panel) and $w_0 \times w_a$ (right panel).}\label{fig:desi_fom}
\end{figure}

In Figures~\ref{fig:ska_fom} and~\ref{fig:desi_fom}, we present the FoM for the SKA and DESI surveys, respectively, as a function of the bin width parameter $\sigma_z$ for three different bin shapes: Gaussian (blue), semi-Gaussian (gray), and top-hat (black dashed). The FoM is computed for the parameter pairs $(H_0, w_0)$ (left panels) and $(w_0, w_a)$ (right panels) following the standard definition. 

For the SKA survey (Fig.~\ref{fig:ska_fom}), the Semi-Gaussian bin shape yields the highest FoM for all $\sigma_z$ values. However, as $\sigma_z$ increases, the FoM of the Gaussian bins drops rapidly. In contrast, top‑hat bins, which have a sharp cut‑off, have the second highest FoM. The Gaussian shaped would obtain the least amount of cosmological information. This indicates that the SKA configuration would find tighter constraints with a Semi-Gaussian configuration. It is also important to note that, one must consider all gain and loss from the bin configuration of choice, top-hat clearly has the smaller shot-noise, but washes out the wiggles easily which is a considerable loss in parameter constraints. The Gaussian binning, despite the accuracy in the $z_c$, has the highest shot-noise.

For the higher‑redshift DESI survey (Fig.~\ref{fig:desi_fom}), the evolution of $D_A(z)$ is slower, so projection smearing is less severe. At small $\sigma_z$, all bins are narrow, and the semi‑Gaussian again performs best due to its smooth window function. However, once $\sigma_z\geq0.035$, the flat top of the semi‑Gaussian begins to loose sensitivity. The Gaussian bin retains a sharper effective redshift distribution and suffers less smearing. The top‑hat bin, despite its uniform weighting, includes the largest number of galaxies (lowest shot noise) and has the best performance. At large $\sigma_z$ the semi‑Gaussian becomes the worst performer, while the top‑hat achieves the highest FoM. This reversal underscores that the optimal bin shape depends not only on the bin width but also on the redshift range and the relative importance of shot noise versus projection effects.

When $\sigma_z$ is large, galaxies with different true redshifts are averaged together within the same bin. 
This coarse binning smears out the characteristic features of the expansion history and structure growth, such as the BAO scale or the shape of the redshift-space distortion pattern. 
Consequently, the cosmological signal is diluted, leading to larger error ellipses and a lower FoM. 
As $\sigma_z$ decreases, the bins become narrower, allowing the analysis to trace the redshift evolution of dark energy more faithfully. 

However, the improvement with smaller $\sigma_z$ is not unbounded. Each redshift bin contains a finite number of galaxies; when bins are made very narrow, the number of objects per bin drops, increasing the shot noise. Therefore, the curves in the figures \ref{fig:ska_fom} and \ref{fig:desi_fom} typically show a steep rise as $\sigma_z$ decreases from large values.

\begin{figure}
    \centering
    \begin{minipage}{0.45\textwidth}
        \centering
        \includegraphics[width=\linewidth]{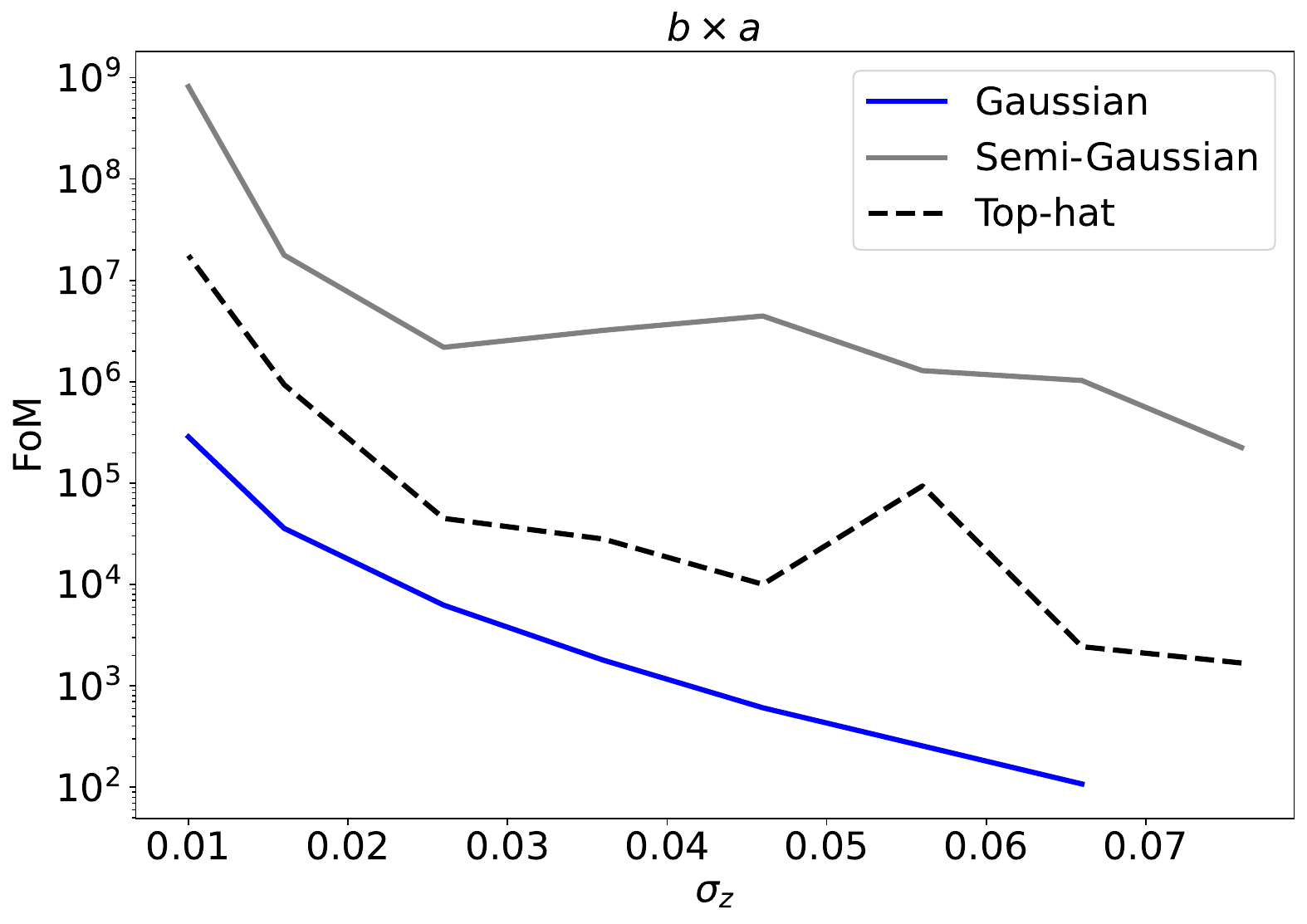}
    \end{minipage}
    \begin{minipage}{0.45\textwidth}
        \centering
        \includegraphics[width=\linewidth]{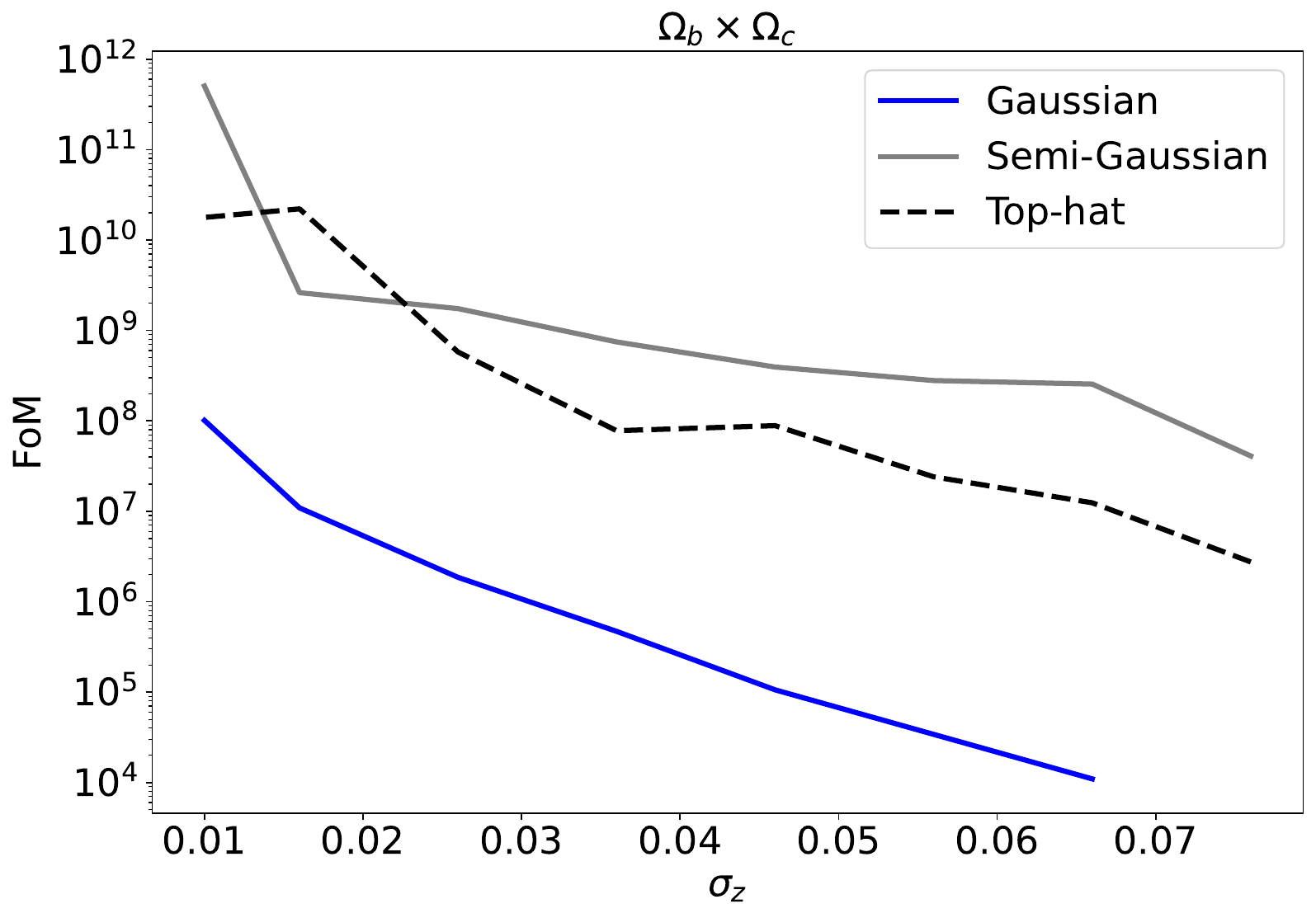}    
    \end{minipage}
    \caption{SKA FoM for different $\sigma_z$ ($\Delta z = 2\sigma_z$) for parameters $b\times a$ (left panel) and $\Omega_b \times \Omega_c$ (right panel).}\label{fig:ska_fom2}
\end{figure}

\begin{figure}
    \centering
    \begin{minipage}{0.45\textwidth}
        \centering
        \includegraphics[width=\linewidth]{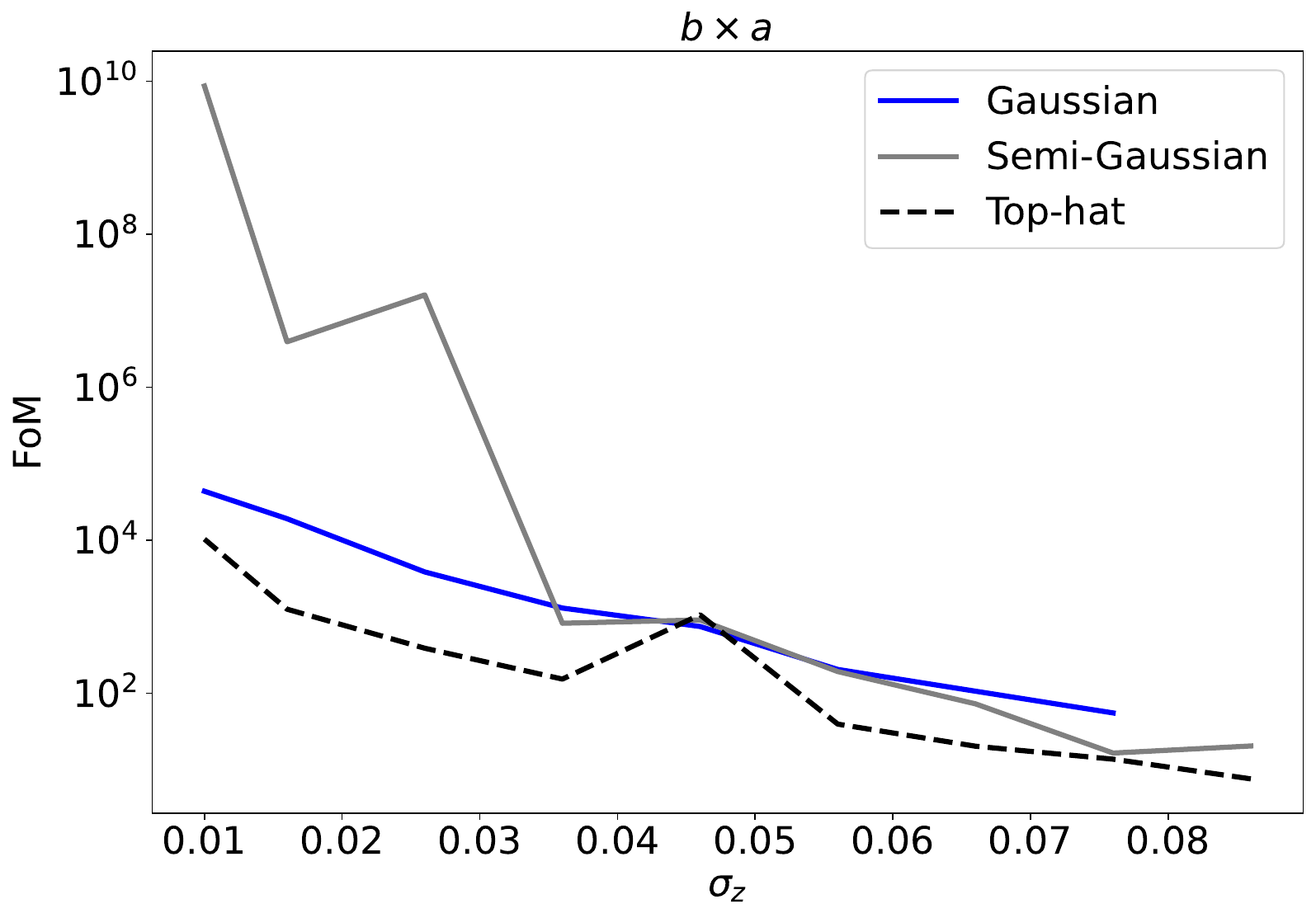}
    \end{minipage}
    \begin{minipage}{0.45\textwidth}
        \centering
        \includegraphics[width=\linewidth]{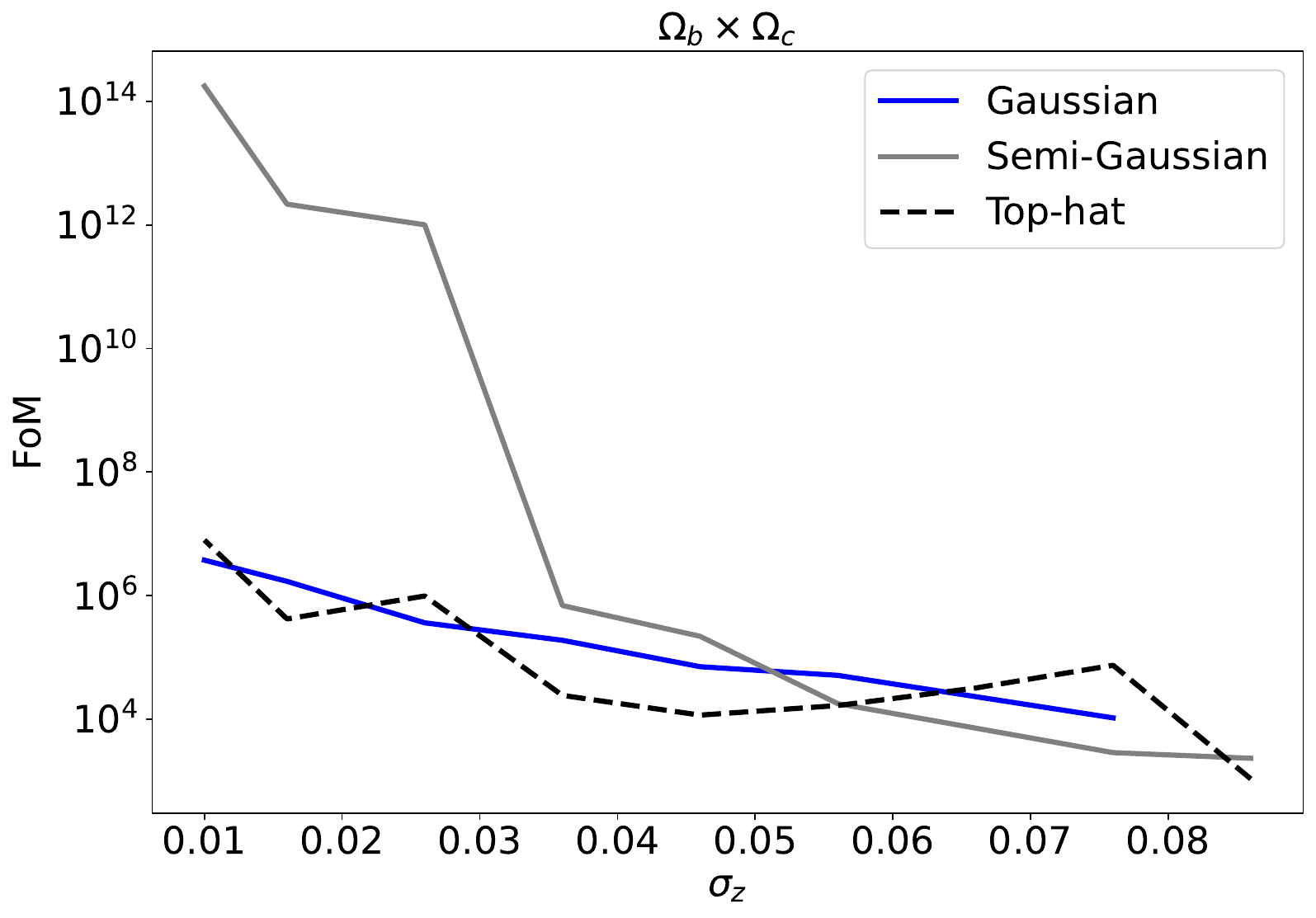}
    \end{minipage}
    \caption{DESI FoM for different $\sigma_z$ ($\Delta z = 2\sigma_z$) for parameters $b\times a$ (left panel) and $\Omega_b \times \Omega_c$ (right panel).}\label{fig:desi_fom2}
\end{figure}

Figures~\ref{fig:ska_fom2} and \ref{fig:desi_fom2} show the FoM for bias and matter related parameters. It is clear that for SKA, semi-Gaussian is the best choice, except for matter related parameter with $\sigma_z\sim 0.015$, where the top-hat separation provides better constraints. DESI has a similar pattern of $H_0$ and dark energy, but the Gaussian bins loose performance in contrast to top-hat, this is due to less number density to provide better information of how galaxies are clustered compared to the total matter. 

In the last section, we found that the highest accuracy for finding the BAO bump from $w(\theta)$ depends on $z_c$ bigger $\sigma_z$ has accurate BAO features, while for tighter constraints using $C_\ell$ the smaller $\sigma_z$ is ideal. For a real data pipeline, one must choose which approach is best: well constrained cosmological parameters, but less accurate BAO measurement; or well measured BAO peak/wiggles. In case one is interested in combining the BAO measurement with other cosmological probes, it might be more interesting to focus on the bin scheme that is statistically robust and with a sharp representation of the redshift of interest, which is the Semi-Gaussian case.

\section{Summary}\label{sec:summary}

The transverse BAO feature is an interesting independent method to constrain cosmological parameters in addition to its 3D counterpart. It is nearly-independent of any fiducial model and can be obtained via $w(\theta)$ or $C_\ell$. However, in cosmology we are interested in the evolution of some observables, in this case, it is ideal to understand the change of the BAO feature in terms of redshift and for that we need to avoid the projection effect that results from the bin's width ($\sigma_z)$. Here, we not only tested the sensitivity to $\sigma_z$, but also tested three alternative bin configurations: Gaussian bins, top-hat, and semi-Gaussian scheme.

Our analysis was based on Fisher Forecast first with $\sigma_z=0.016$ for the SKA HI and DESI LRG surveys, lower and higher redshift samples, respectively. In the SKA first result showed that the semi-Gaussian scenario constrains parameters better than the top-hat, despite having a bigger shot-noise. However, for the DESI example, the top-hat performed better, but with comparable constraints. This was the first indication that the semi-Gaussian is an accurate bin configuration.

We tested the accuracy of finding $\theta_{\rm BAO}(z_c)$ for a series of $\sigma_z$ of the three types of bins. We compared the simulated result with the expected result from fiducial cosmology by $\alpha$. We found that in the lower redshift case the semi-Gaussian bin has the most accurate BAO position of the three configurations, and the top-hat showed big discrepancies from the expected $\theta_{\rm BAO}$. This proves to be a strong indicator for the community to avoid relying on top-hat bin configurations in the nearly-local universe because the BAO feature is deeply affected by the projection effect.

For the DESI $\alpha$ tests, we noticed an improvement in finding the BAO feature for a higher $\sigma_z$, as expected for a higher redshift survey. In this case, semi-Gaussian and Gaussian are comparable. The dependence on $\sigma_z$ and the parameter constraints was analysed using Fisher Forecast. We found that in most cases the semi-Gaussian bin separation increases the precision of $H_0$, $w_0$, and $w_a$ for all $\sigma_z$ tested in the SKA configuration. However, for DESI, once $\sigma_z\geq0.035$, the semi‑Gaussian begins to loose information. We conclude that the choice of bin configuration that effectively represents the $z_c$, which may be intertwined with systematics, is the most reliable solution to avoid projection effects in the transverse BAO analysis for small redshift surveys, but for higher redshift surveys the bigger $\sigma_z$ clearly looses cosmological precision due to the smearing of the BAO signal. 

Using our proposed method to correct for the projection effect, we observe that in both configurations the projection effect can be reduced. However, for $\sigma_z>0.05$, it becomes impossible to recover information suitable for an accurate BAO measurement. 

A natural concern, which we anticipate from the perspective of precision cosmology, is whether the transverse BAO can ever achieve percent-level 
accuracy, given the large projection effects we have identified. Our forecast provides a clear and optimistic answer: yes, provided that the binning is chosen 
carefully.

The central warning of this work is that, in the absence of careful bin optimisation, projection effects can bias the recovered BAO scale by 10--50\% or 
more. This is not a statistical fluctuation but a systematic error arising from the mixing of BAO signals from different epochs within a finite redshift bin. 
The effect is most severe for low-redshift surveys, for broad bins ($\sigma_z > 0.04$), and for top-hat bin shapes. However, we have also 
demonstrated that these biases can be controlled. Our proposed correction ($\tilde{\alpha}$), which uses adjacent redshift slices to statistically remove the 
projection bias, recovers the BAO scale to percent-level accuracy for $\sigma_z \lesssim 0.04$, as shown in Figures~\ref{fig:alphac1} and~\ref{fig:alphac2}. For larger 
$\sigma_z$, the correction becomes less efficient, indicating that bin widths should be kept small.

Among the three bin shapes tested, the semi-Gaussian binning offers the most robust performance. It provides the best compromise between low shot noise 
(via its flat top) and accurate representation of the central redshift (via its Gaussian wings). For low-redshift surveys, it yields the lowest failure rates 
and the most accurate $\alpha$ values. For higher-redshift surveys, it remains competitive with or outperforms the alternatives up to $\sigma_z \lesssim 0.04$.

In practice, percent-level transverse BAO measurements can be achieved by combining three elements: semi-Gaussian binning, the projection correction 
described in Section~\ref{alpha}, and mock-based calibration to account for residual systematics. Once these controls are in place, the transverse BAO provides a 
model-independent measurement of the angular diameter distance $D_A(z)$ that is complementary to the 3D BAO. This is particularly valuable for breaking 
degeneracies in dark energy parameter constraints, cross-checking results from spectroscopic surveys, and exploiting the vast statistical power of photometric 
surveys where 3D BAO is not feasible.

In conclusion, our work demonstrates that the transverse BAO is not inherently limited by projection effects; rather, the limitation lies in the choice of 
binning. Several points need to be considered before employing a spectroscopic galaxy survey to search for $\theta_{\rm BAO}$. First, the redshift range, a relatively local universe is susceptible to non-linearities which cause problems for mock construction and model estimation. In terms of bin selection, low redshift surveys are easily affected by the projection effect due to FoG. Thus, for low-redshift surveys, semi-Gaussian binning provides a more suitable slicing. For higher redshifts, it still yields tighter parameter constraints than alternative schemes up to a certain $\sigma_z$. For a real data pipeline, one must choose which approach is best: well constrained cosmological parameters, but less accurate BAO measurement; or well measured BAO peak/wiggles. In case one is interested in combining the BAO measurement with other cosmological probes, it might be more interesting to focus on the bin scheme that is statistically robust and with a sharp representation of the redshift of interest, which is the Semi-Gaussian case.

\newpage
\acknowledgments
PSF thanks Brazilian funding agency CNPq for financial support PCI-DB  316199/2025-7. CB acknowledges financial support from the CNPq grant 306630/2025-7. 
This work made use of the Milliways HPC computer located at the Instituto de Física in the Universidade Federal do Rio de Janeiro, managed and funded by \href{https://sites.google.com/view/arcos-ufrj/about?authuser=0}{ARCOS} (Astrophysics, Relativity and COSmology research group). This work was developed thanks to the National Observatory Data Center (CPDON). The authors are grateful to the anonymous referee for valuable comments and constructive suggestions.

\bibliographystyle{JHEP}
\bibliography{biblio.bib}

\appendix
\section{Covariance matrix}\label{ap:covs}
\begin{figure}
\centering
\includegraphics[width=0.8\textwidth]{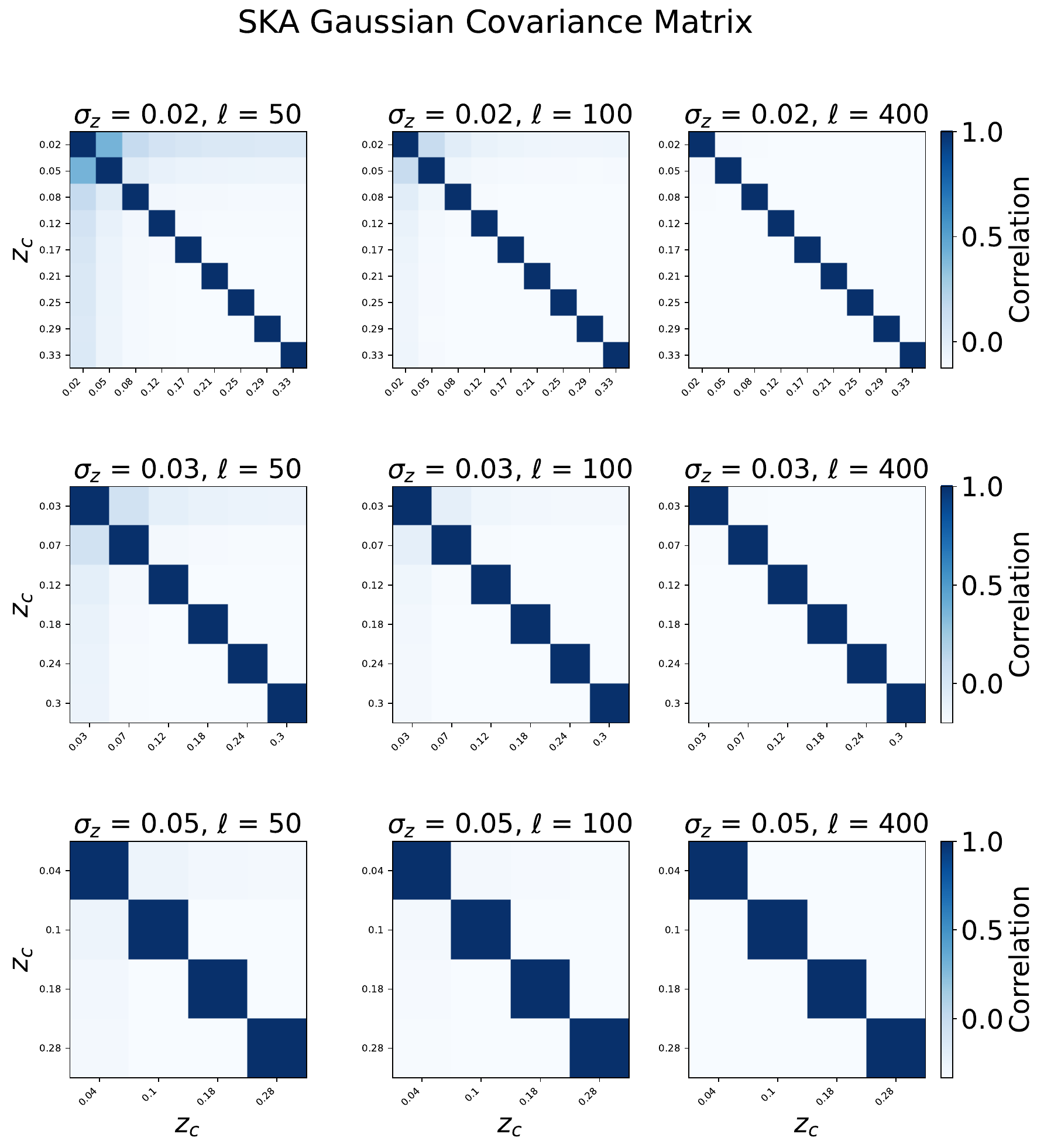}
\caption{Covariance matrices for the SKA survey using Gaussian redshift bins. Each subplot shows the correlation between bin centres $z_c$ for a fixed input redshift $z$ (rows) and bin width $\Delta z$ (columns). The colour bar indicates the correlation coefficient.}
\label{fig:ska_g}
\end{figure}
Figure~\ref{fig:ska_g} displays the correlation matrices obtained for the SKA survey when a Gaussian bin shape is adopted. The three rows correspond to input redshifts $z = 0.02$, $0.03$ and $0.05$, while the three columns represent the maximum multipole moment $\ell$ used in the Fisher forecast ($\ell_{\mathrm{max}}=50$, $100$, $400$). For $\ell=50$, only large angular scales contribute; the off‑diagonal correlations are moderate, reflecting the relatively smooth window functions. As $\ell$ increases to $100$ and $400$, smaller scales are included, and the covariance becomes more diagonally dominant because fine‑scale modes are more independent. These trends illustrate how the choice of $\ell_{\mathrm{max}}$ affects the covariance structure in BAO analyses.
\begin{figure}
\centering
\includegraphics[width=0.8\textwidth]{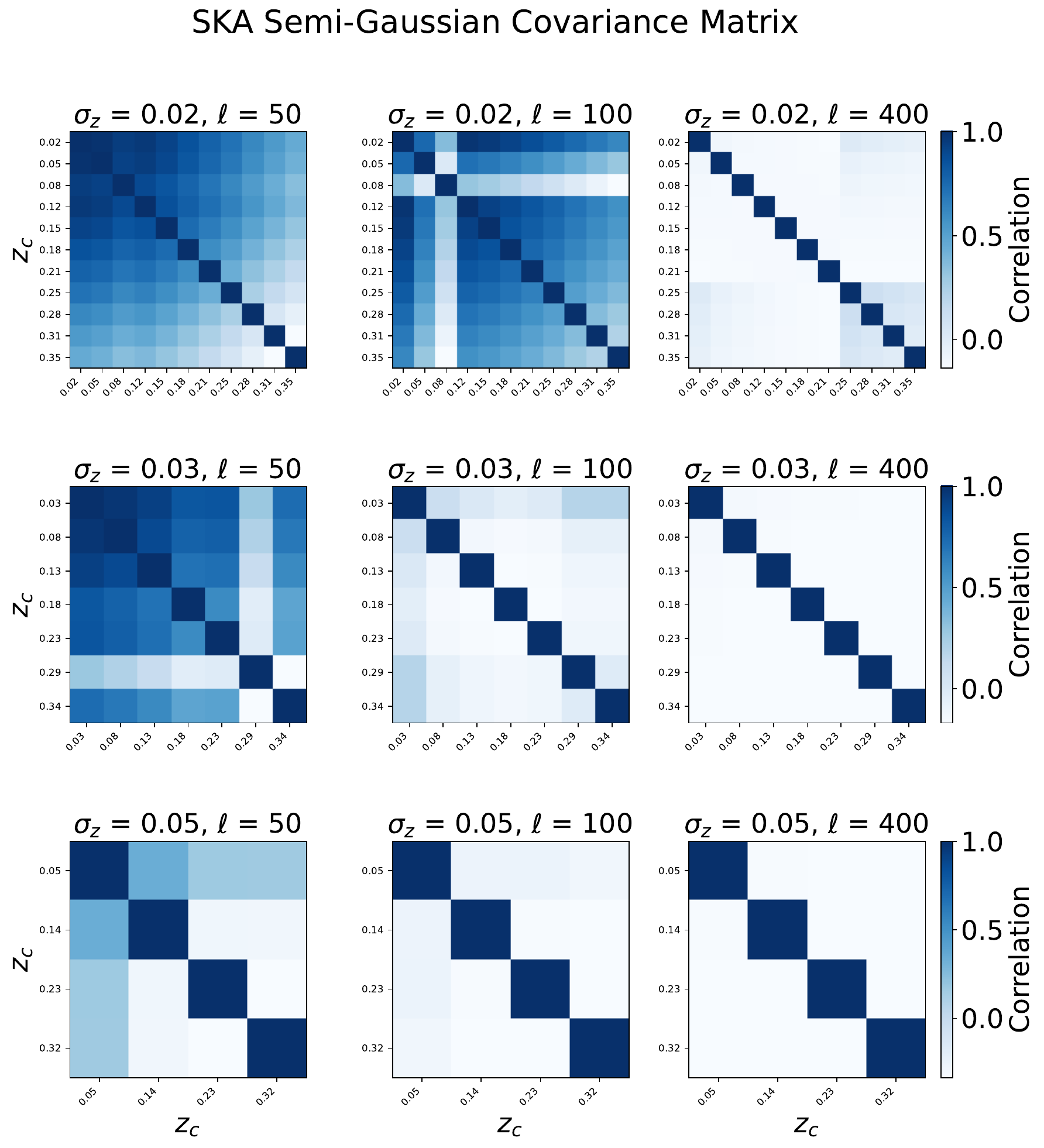}
\caption{Same as Fig.~\ref{fig:ska_g} but for semi‑Gaussian redshift bins. The semi‑Gaussian shape is a hybrid that retains a flat top and Gaussian wings.}
\label{fig:ska_sg}
\end{figure}

The semi‑Gaussian bin shape (Fig.~\ref{fig:ska_sg}) produces covariance matrices qualitatively similar to the pure Gaussian case but with slightly stronger off‑diagonal correlations for the same $\ell$ and redshift. The flat top increases overlap between adjacent bins, especially at lower $\ell$ where large scales dominate. For $\ell=50$, the correlation differences are most pronounced at $\sigma_z=0.02$, with the semi‑Gaussian showing broader tails. 

\begin{figure}
\centering
\includegraphics[width=0.8\textwidth]{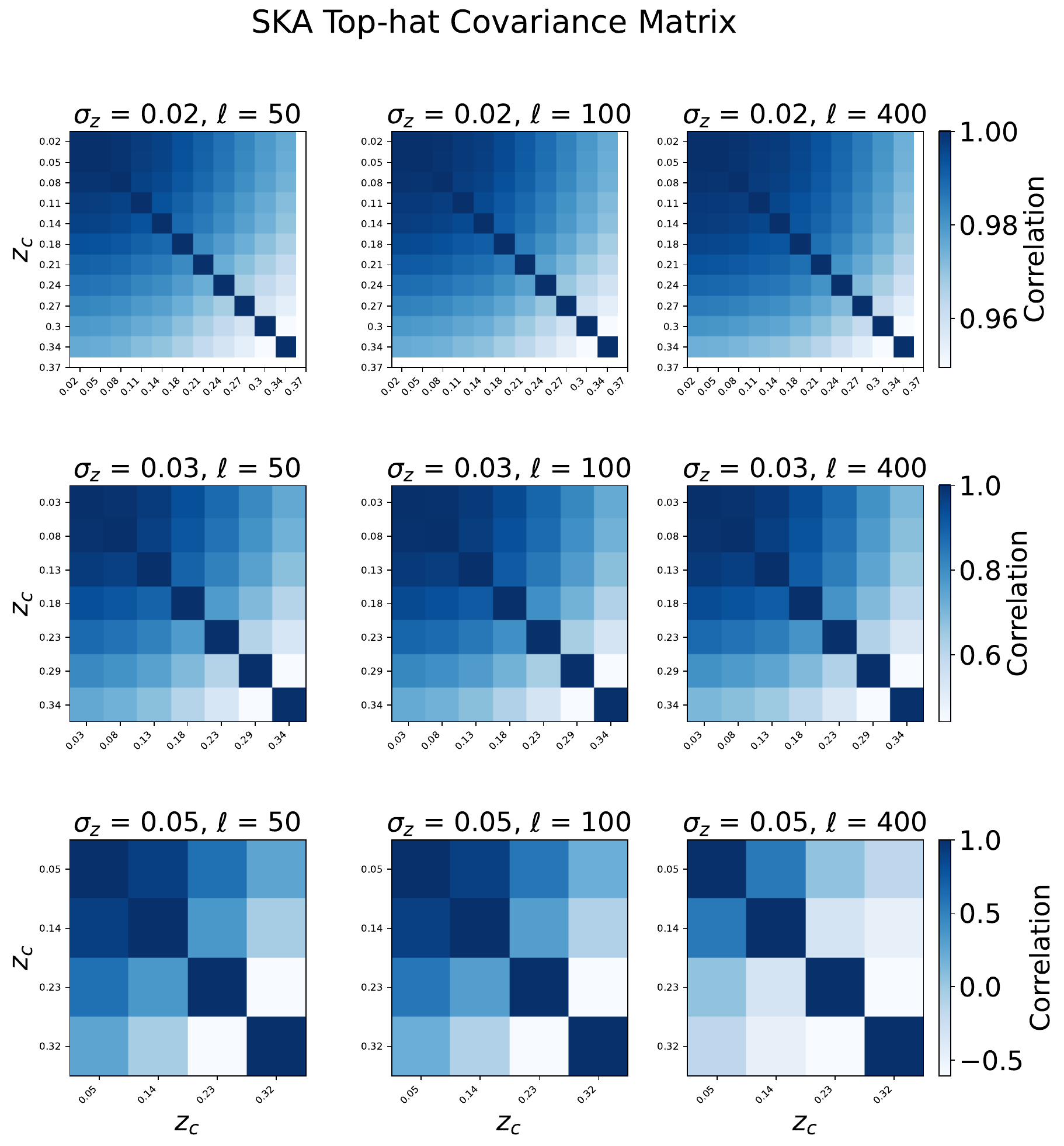}
\caption{Correlation matrices for the SKA survey using top‑hat (boxcar) redshift bins. The colour scale is adjusted to highlight the very high correlations (note the lower bound at $0.5$ in the right‑most panels).}
\label{fig:ska_th}
\end{figure}

Figure~\ref{fig:ska_th} presents the covariance matrices for top‑hat bins. Unlike smooth Gaussian or semi‑Gaussian shapes, the top‑hat window function has sharp edges, leading to near‑unit correlations between bins whose centres are close, even at high $\ell$. For $\ell=50$, adjacent bins exhibit correlations above $0.96$, indicating almost complete redundancy. As $\ell$ increases to $400$, the correlations remain above $0.5$ for bin separations up to $\sim0.1$ in $z_c$. Such high correlations imply that the number of independent redshift slices is greatly reduced, regardless of $\ell_{\mathrm{max}}$. Consequently, while top‑hat bins may simplify some aspects of the analysis, they are less efficient for extracting transverse BAO information.
\begin{figure}
\centering
\includegraphics[width=0.8\textwidth]{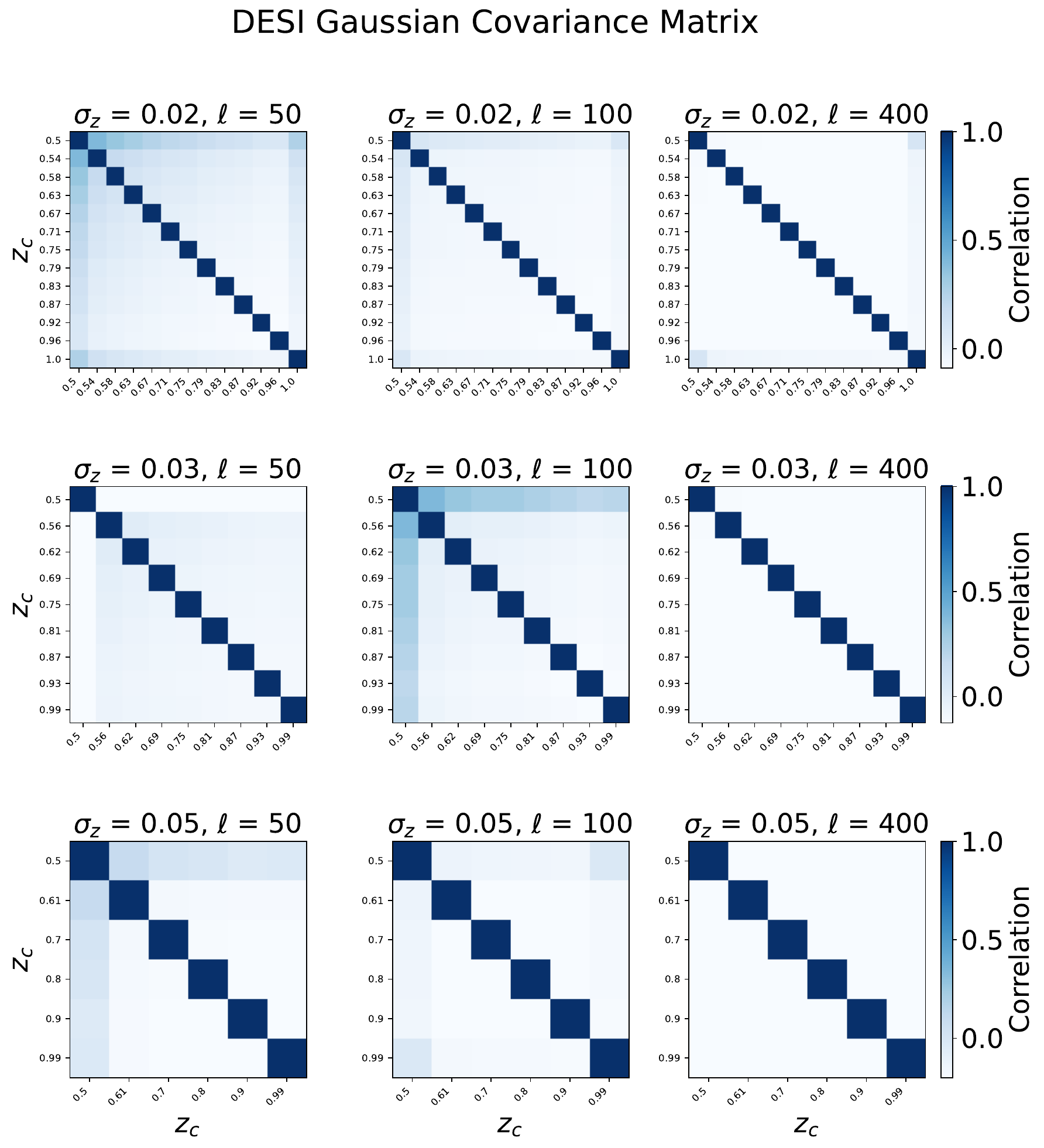}
\caption{Same as Fig.~\ref{fig:ska_g} but for the DESI survey (redshift range $z_c \approx 0.5$–$1.0$).}
\label{fig:desi_g}
\end{figure}

The DESI survey covers higher redshifts (Fig.~\ref{fig:desi_g}) with $z_c$ ranging from $0.5$ to $1.03$. The Gaussian bin shape again yields moderate off‑diagonal correlations that decrease with increasing $\ell$. For $\ell=50$, correlations are relatively broad, but they become increasingly diagonal as $\ell$ reaches $400$. Compared to SKA at similar $\ell$, DESI correlations are slightly lower for the same redshift, likely due to the higher redshift where the comoving radial distance changes more slowly with $z$, reducing mode mixing. This makes DESI more resilient to bin‑induced correlations, which is advantageous for BAO measurements.
\begin{figure}
\centering
\includegraphics[width=0.8\textwidth]{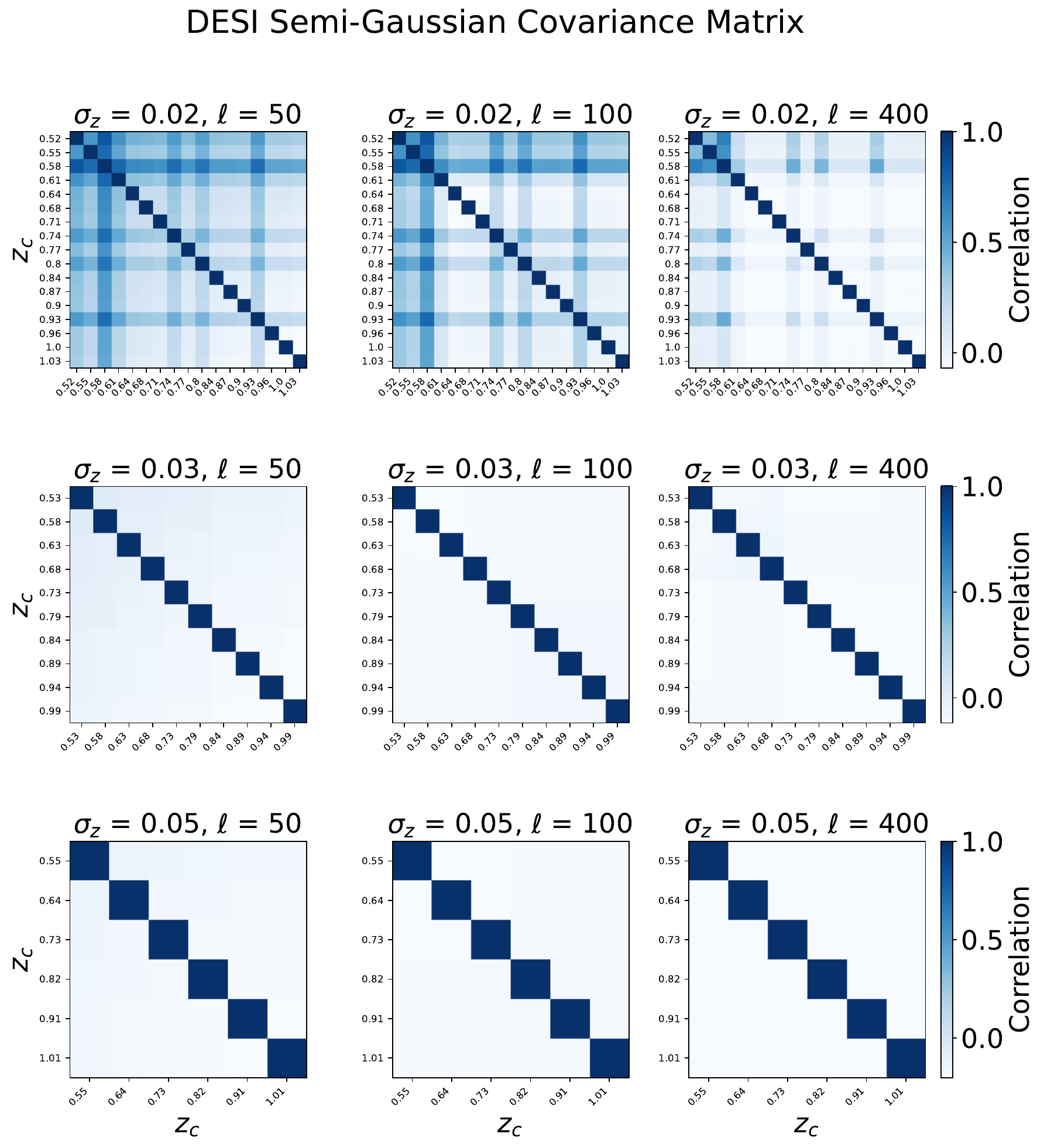}
\caption{Same as Fig.~\ref{fig:ska_sg} but for the DESI survey.}
\label{fig:desi_sg}
\end{figure}

The semi‑Gaussian bins for DESI (Fig.~\ref{fig:desi_sg}) exhibit correlations that lie between the pure Gaussian and top‑hat cases. For $\ell=50$, off‑diagonal elements are only slightly larger than those of the Gaussian, but at $\ell=400$, correlations of $\sim0.7$ persist over a redshift separation of $\sim0.1$. The flat top ensures that bins centred within the flat region are very strongly correlated at low $\ell$, while the Gaussian wings cause a smoother fall‑off at higher $\ell$. 
\begin{figure}
\centering
\includegraphics[width=0.8\textwidth]{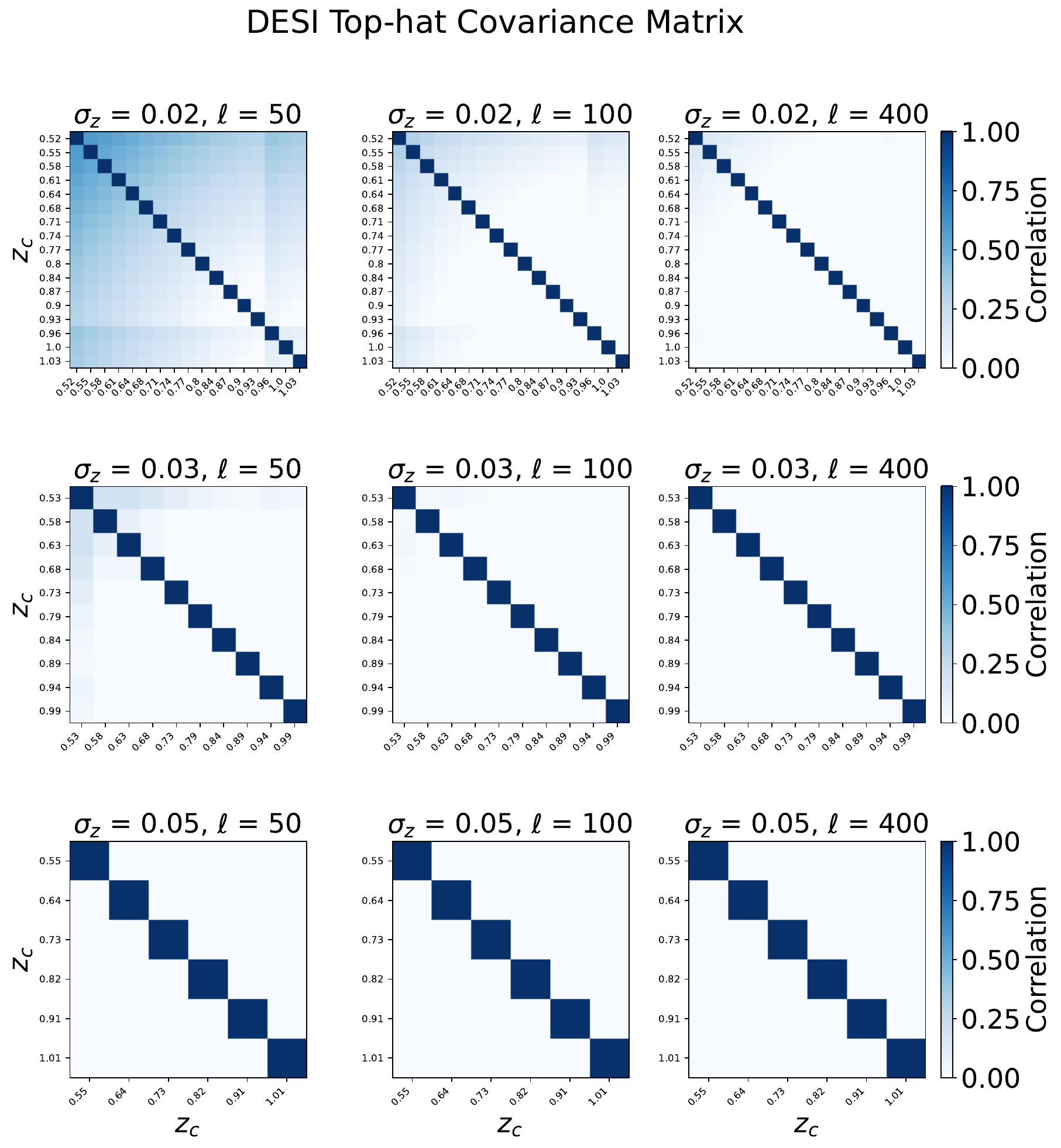}
\caption{Same as Fig.~\ref{fig:ska_th} but for the DESI survey. The correlation scale is adjusted to show the high values typical of top‑hat bins.}
\label{fig:desi_th}
\end{figure}
The final set of matrices (Fig.~\ref{fig:desi_th}) corresponds to top‑hat bins for DESI. As with SKA, the correlations are higher for lower $\ell$.

We emphasise that the binning schemes presented here are constructed as disjoint partitions of the galaxy catalogue: each galaxy is assigned to exactly one bin according to the maximum of the selection function. The window functions shown in Figures~2 and~3 are used only to project the power spectrum; they do not imply that galaxies are shared between bins. Consequently, the covariance between bins arises from projection effects and the finite sky coverage, not 
from overlapping galaxy samples.

The covariance matrix in our analysis is computed in harmonic ($\ell$) space, where the correlation between bins arises from the cross-power spectra $C_\ell^{i,j}$ . Although the galaxy populations in adjacent bins are disjoint, the underlying matter power spectrum projected into each bin still produces non-zero correlations because the radial kernels have finite support. The magnitude of these correlations depends on the shape and smoothness of the window functions. Smooth Gaussian kernels produce localised radial transfer functions, yielding a more diagonal covariance matrix. In contrast, the sharp edges of top-hat bins introduce oscillatory sidelobes in the Fourier transform, producing long-range correlations in harmonic space even when the 
bins themselves are disjoint in redshift assignment. The semi-Gaussian window exhibits intermediate behaviour: the flat top broadens the kernel and increases correlations relative to Gaussian, while the smooth wings suppress the strongest oscillatory features of a pure top-hat.

This pattern is more pronounced for the SKA survey than for DESI. At low 
redshifts, the angular diameter distance evolves more rapidly with redshift, 
leading to stronger mixing of BAO signals from different epochs within a finite 
bin. Consequently, the differences between bin shapes—particularly the 
superiority of Gaussian binning in producing a diagonal covariance—are amplified 
for SKA. For DESI, which probes higher redshifts where the evolution of the 
angular diameter distance is slower, the covariance matrices are generally more 
diagonal across all bin shapes, and the differences between the three schemes 
are less dramatic. This highlights that the choice of binning is particularly 
critical for low-redshift surveys, where projection effects are most severe.

\section{BAO angular scale recovery}\label{ap:b}

In this appendix, we present supplementary figures showing the recovered BAO angular scale $\alpha \equiv \theta_{\mathrm{fit}} / \theta_{\mathrm{BAO}}^{\mathrm{fid}}$ for each bin configuration and survey. As described in Section~\ref{sec:alpha}, $\alpha$ is obtained using a hybrid model-free procedure that combines information from the angular power spectrum $C_\ell$ and the angular correlation function $w(\theta)$. 

\begin{figure}
\centering
\includegraphics[width=.7\textwidth]{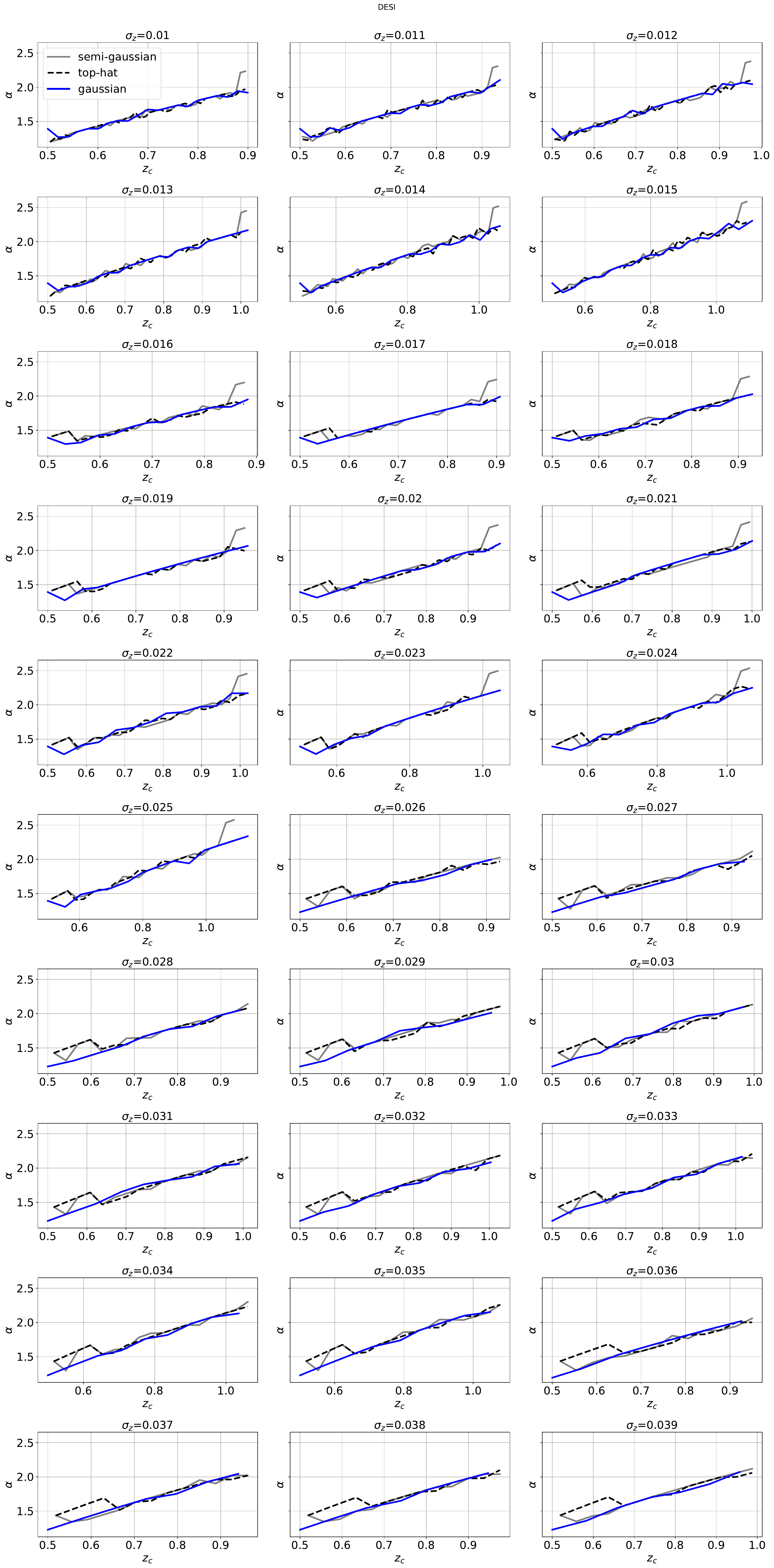}
\caption{Recovered BAO angular scale $\alpha$ as a function of bin centre $z_c$ for the DESI-like survey. Each panel shows a different bin width for the three configurations. The blue solid line represents the Gaussian bin separation, while the dashed black line represents the top-hat one, and the gray solid line the semi-Gaussian bin pattern.}
\label{fig:app:desi_alpha}
\end{figure}

\begin{figure}
\centering
\includegraphics[width=.7\textwidth]{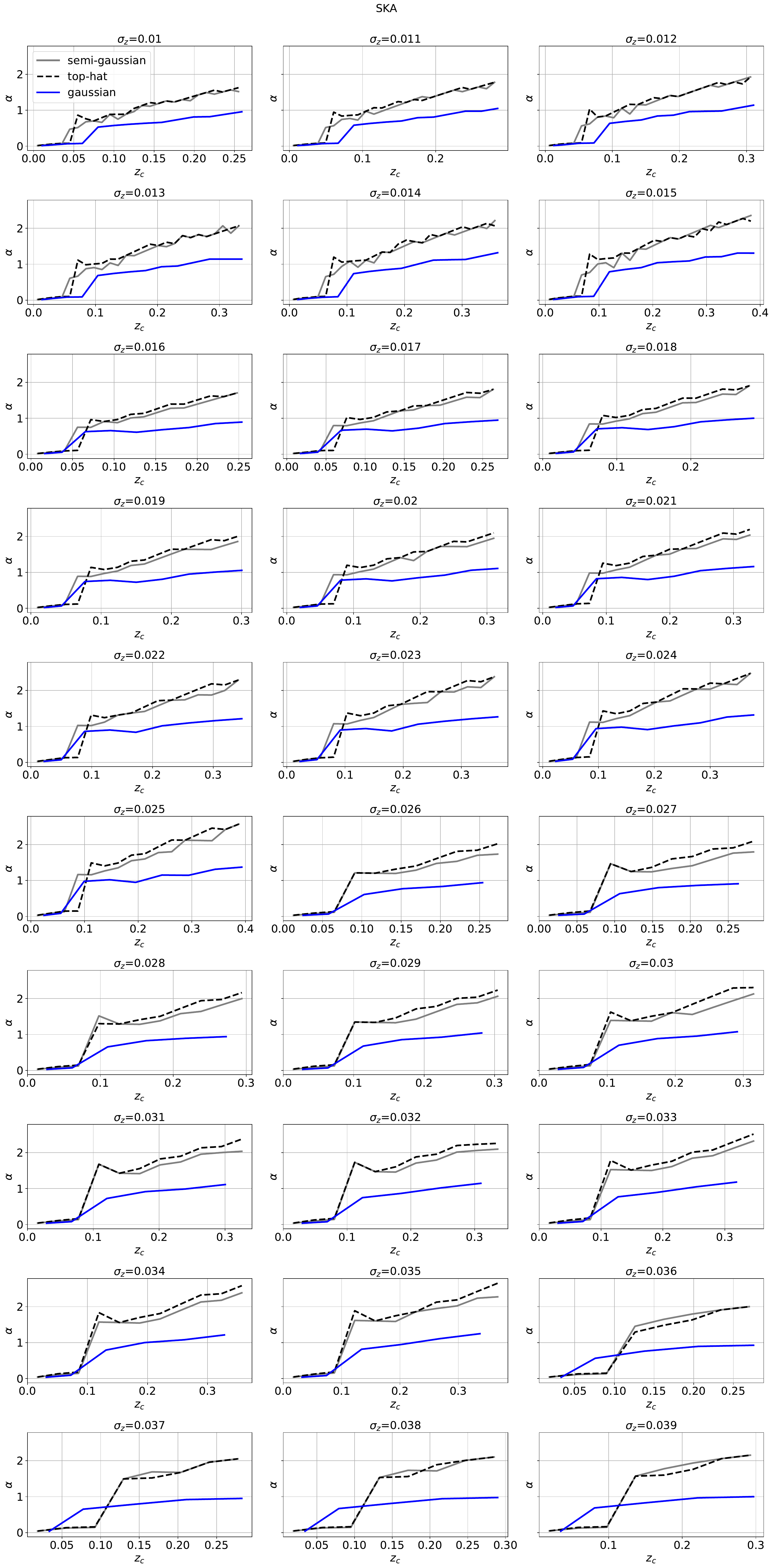}
\caption{Same as Figure~\ref{fig:app:desi_alpha}, but for the SKA-like survey, which probes lower redshifts ($z_c \lesssim 0.3$) where projection effects are more severe.}
\label{fig:app:ska_alpha}
\end{figure}

Figures~\ref{fig:app:desi_alpha} and~\ref{fig:app:ska_alpha} show $\alpha$ as a function of the bin centre $z_c$ for the DESI and SKA surveys, respectively. Each panel corresponds to a different bin shape (Gaussian, semi-Gaussian, and top-hat), and the different coloured curves represent different bin widths $\sigma_z$, ranging from $0.01$ to $0.04$ in steps of $0.001$. The curves are colour-coded by $\sigma_z$, with lighter colours indicating narrower bins. Only cases where the BAO detection passed the failure criterion (defined in Section~\ref{sec:alpha}) are shown.

For the DESI survey (Figure~\ref{fig:app:desi_alpha}), the recovered $\alpha$ values are generally close to unity for narrow bins ($\sigma_z \lesssim 0.025$), with scatter increasing as the bin width grows. The Gaussian and semi-Gaussian bin shapes show similar behaviour, with $\alpha$ remaining within approximately 5\% of unity for $\sigma_z \lesssim 0.03$. The top-hat bins exhibit larger deviations, particularly at the edges of the redshift range and for broader bins, consistent with the sharp edges introducing oscillatory features in the radial kernel.

For the SKA survey (Figure~\ref{fig:app:ska_alpha}), the sensitivity to binning choices is much stronger. The SKA probes lower redshifts, where the angular diameter distance evolves more rapidly with redshift, making the BAO feature more susceptible to projection effects. For Gaussian bins, $\alpha$ deviates from unity by up to 50\% for broader bins, with deviations becoming more pronounced at higher $z_c$ within the SKA range. The semi-Gaussian bins show markedly better performance: $\alpha$ remains close to unity across all $z_c$ and $\sigma_z$, with deviations typically below 10\% even for the broadest bins. This demonstrates the advantage of the semi-Gaussian shape, which combines the low shot noise of a flat top with the smooth localisation of Gaussian wings. The top-hat bins, by contrast, exhibit the largest scatter and the most severe deviations.

\begin{figure}
\centering
\includegraphics[width=.8\textwidth]{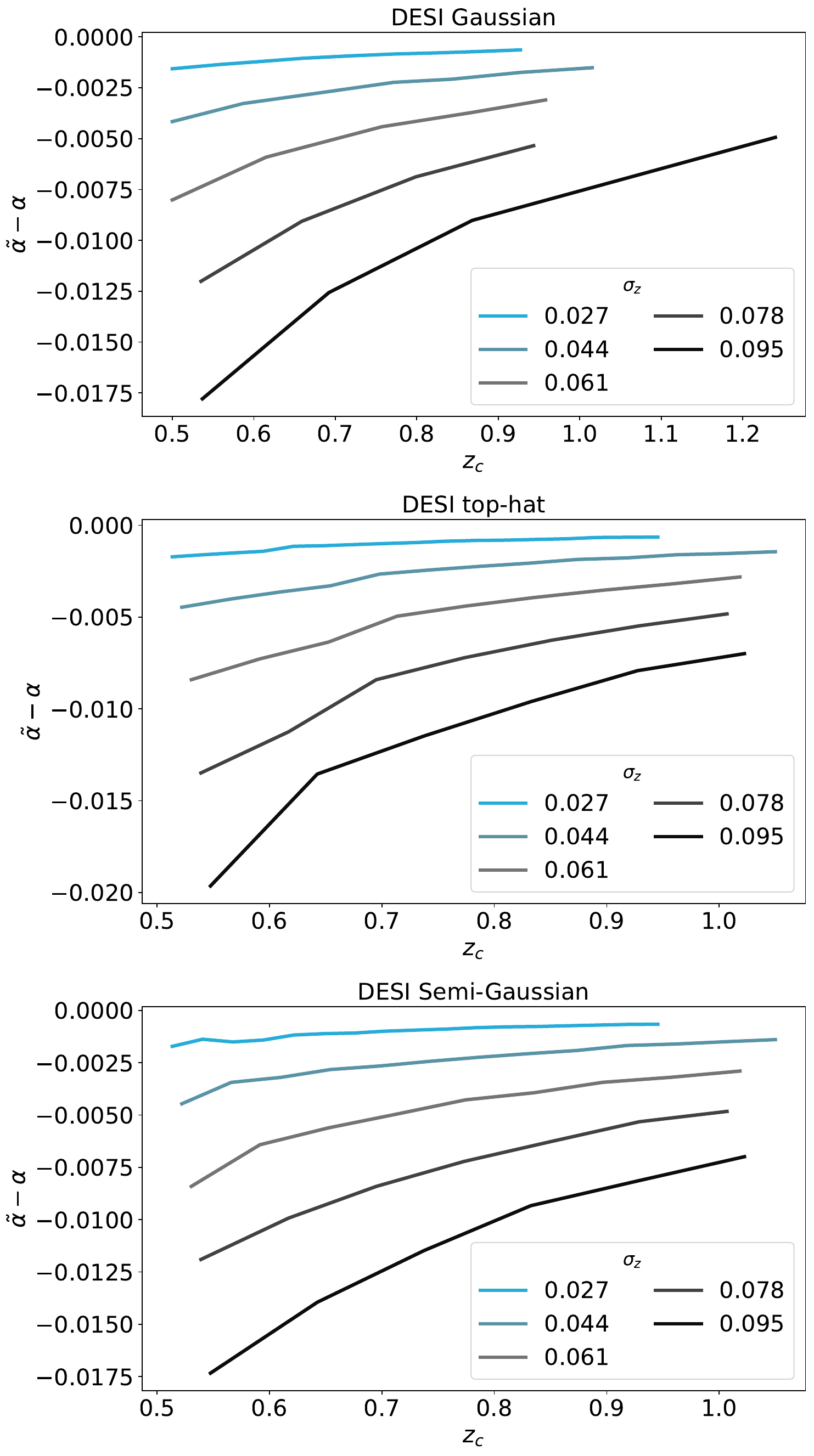}
\caption{Projection correction $\tilde{\alpha} - \alpha$ as a function of bin centre $z_c$ for the DESI-like survey. The correction is largest for broad bins and top-hat shapes, but remains small for $\sigma_z \lesssim 0.04$.}
\label{fig:app:desi_correction}
\end{figure}

\begin{figure}
\centering
\includegraphics[width=.8\textwidth]{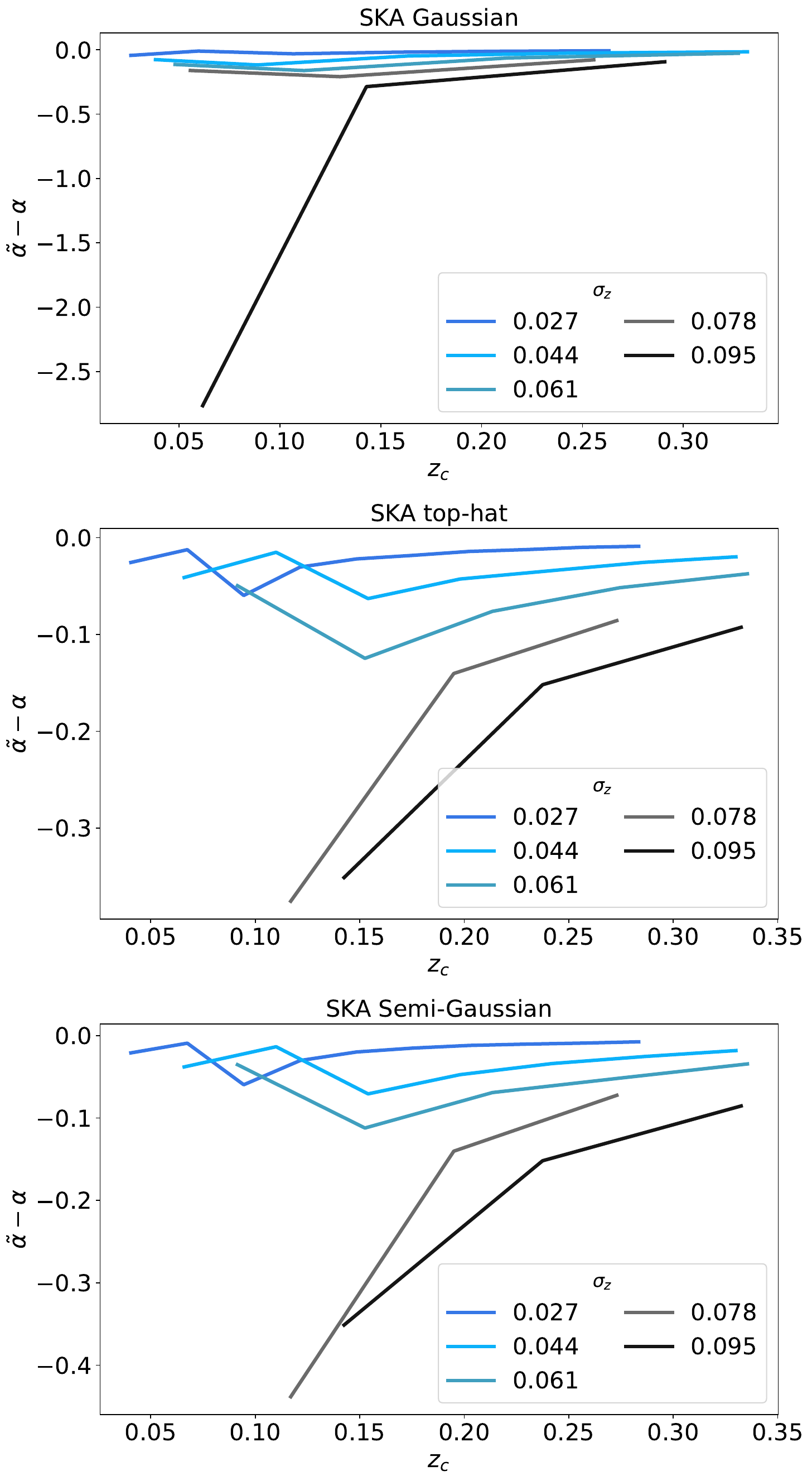}
\caption{Same as Figure~\ref{fig:app:desi_correction}, but for the SKA-like survey. The correction is more substantial at low redshifts, reflecting the stronger projection effects in the local Universe.}
\label{fig:app:ska_correction}
\end{figure}

Figures~\ref{fig:app:desi_correction} and~\ref{fig:app:ska_correction} show the projection correction $\tilde{\alpha} - \alpha$ as a function of $z_c$. These plots demonstrate the effectiveness of our correction scheme: for $\sigma_z \lesssim 0.04$, the correction brings $\alpha$ back to within percent-level accuracy of unity. For broader bins, the correction becomes less efficient, as expected. The correction is more substantial at low redshifts, reflecting the stronger projection effects in the local Universe.

In summary, the results presented in this appendix reinforce the main conclusions of the paper: the semi-Gaussian binning provides the most robust performance across a wide range of $\sigma_z$, particularly for low-redshift surveys. The correction method proposed in Section~\ref{sec:alpha} can recover percent-level accuracy for bins with $\sigma_z \lesssim 0.04$, but fails for broader bins, underscoring the importance of keeping bin widths sufficiently small.

\end{document}